\documentclass[twocolumn]{aastex63}

\usepackage{amssymb}
\usepackage{amsmath}

\usepackage{url}
\usepackage[flushleft]{threeparttable}
\received{XXX}
\revised{YYY}
\accepted{ZZZ}
\submitjournal{ApJ}

\shorttitle{DWFS II.: Photometry of Massive Galaxies}
\shortauthors{Miller et al.}
\graphicspath{{./}{figures/}}

\begin{document}

\title{The Dragonfly Wide Field Survey. II. Accurate Total Luminosities and Colors of Nearby Massive Galaxies and Implications for the Galaxy Stellar Mass Function}

\correspondingauthor{Tim B. Miller}
\email{tim.miller@yale.edu}

\author[0000-0001-8367-6265]{Tim B. Miller}
\affiliation{Department of Astronomy, Yale University, New Haven, CT 06511, USA \\}

\author[0000-0002-8282-9888]{Pieter van Dokkum}
\affiliation{Department of Astronomy, Yale University, New Haven, CT 06511, USA \\}

\author[0000-0002-1841-2252]{Shany Danieli}
\altaffiliation{NASA Hubble Fellow}
\affil{Department of Physics, Yale University, New Haven, CT 06520, USA \\}
\affil{Yale Center for Astronomy and Astrophysics, Yale University, New Haven, CT 06511, USA \\}
\affil{Department of Astronomy, Yale University, New Haven, CT 06511, USA \\}
\affil{Institute for Advanced Study, 1 Einstein Drive, Princeton, NJ 08540, USA}

\author[0000-0001-9592-4190]{Jiaxuan Li}
\affiliation{Kavli Institute for Astronomy and Astrophysics, Peking University, 5 Yiheyuan Road, Haidian District, Beijing 100871, China\\}

\author[0000-0002-4542-921X]{Roberto Abraham}
\affiliation{Department of Astronomy \& Astrophysics, University of Toronto, 50 St. George Street, Toronto, ON M5S 3H4, Canada\\}
\affiliation{Dunlap Institute for Astronomy and Astrophysics, University of Toronto, Toronto ON, M5S 3H4, Canada\\}

\author[0000-0002-1590-8551]{Charlie Conroy}
\affiliation{Harvard-Smithsonian Center for Astrophysics, 60 Garden Street, Cambridge, MA 02138, USA\\}

\author[0000-0002-8931-4684]{Colleen Gilhuly}
\affiliation{Department of Astronomy \& Astrophysics, University of Toronto, 50 St. George Street, Toronto, ON M5S 3H4, Canada\\}

\author[0000-0003-4970-2874]{Johnny P. Greco}
\altaffiliation{NSF Astronomy \& Astrophysics Postdoctoral Fellow}
\affiliation{Center for Cosmology and AstroParticle Physics (CCAPP), The Ohio State University, Columbus, OH 43210, USA}

\author[0000-0002-7490-5991]{Qing Liu}
\affiliation{Department of Astronomy \& Astrophysics, University of Toronto, 50 St. George Street, Toronto, ON M5S 3H4, Canada\\}

\author[0000-0002-2406-7344]{Deborah Lokhorst}
\affiliation{Department of Astronomy \& Astrophysics, University of Toronto, 50 St. George Street, Toronto, ON M5S 3H4, Canada\\}
\affiliation{Dunlap Institute for Astronomy and Astrophysics, University of Toronto, Toronto ON, M5S 3H4, Canada\\}

\author[0000-0001-9467-7298]{Allison Merritt}
\affiliation{Max-Planck-Institut f¨ur Astronomie, K¨unigstuhl 17, D-69117 Heidelberg, Germany}

\begin{abstract}
Stellar mass estimates of massive galaxies are susceptible to systematic errors in their photometry, due to their extended light profiles. In this study we use data from the Dragonfly Wide Field Survey (DWFS) to accurately measure the total luminosities and colors of nearby massive galaxies. The low surface brightness limits of the survey ($\mu_g \approx $ 31 mag arcsec $^{-2}$ on a one arcmin scale) allows us to implement a method, based on integrating the 1-D surface brightness profile, that is minimally dependent on any parameterization. We construct a sample of 1188 massive galaxies with $\log M_*/M_\odot > 10.75$ based on the Galaxy Mass and Assembly (GAMA) survey and measure their total luminosities and $g-r$ colors. We then compare our measurements to various established methods applied to imaging from the Sloan Digital Sky Survey (SDSS), focusing on those favored by the GAMA survey. In general, we find that galaxies are brighter in the $r$ band by an average of ${\sim}0.05$ mag and bluer in $g-r$ colors by $\sim 0.06$ mag compared to the GAMA measurements. These two differences have opposite effects on the stellar mass estimates. The total luminosities are larger by $5\%$ but the mass-to-light ratios are lower by $\sim 10\%$. The combined effect is that the stellar mass estimate of massive galaxies decreases by $7\%$. This, in turn, implies a small change in number density of massive galaxies: $\leq 30\%$ at $\log M_*/M_\odot \geq 11$. 
\end{abstract}


\section{Introduction} 
\label{sec:intro}
The total stellar mass of a galaxy represents the combined result of all the physical processes which affect its formation and evolution. Thus, the stellar mass function of galaxies (hereafter, SMF) represents one of the fundamental probes of galaxy formation and an important constraint on theoretical models. It is well known that the SMF cuts off at $ \log M_*/M_\odot \gtrsim 11$, with the exponential decline at higher masses often attributed to feedback by active galactic nuclei (AGN) \citep{Blumenthal1984,Birnboim2003,Keres2005,Dekel2006,Catteneo2006}

Due to its exponential decline, the high mass end of the SMF is extremely sensitive to systematic biases in the calculation of stellar masses. A systematic effect on the order of 10\% in the calculation of stellar masses can lead to a factor of 2 difference in the number density of massive galaxies. To measure the slope of the exponential decline, and to understand its physical implications, it is crucial to accurately measure the total stellar masses contained in massive galaxies. Subtle differences have led to disagreement in the literature about the slope of the high mass end of the SMF \citep{Bell2003,Li2009,Moustakas2013,Bernardi2013, Dsouza2015,Thanjavur2016,Bernardi2017,Wright2017, Kravtsov2018}

Determining the stellar mass of an individual massive galaxy is difficult, and one of the main challenges is determining the galaxy's total flux. Photometric techniques are often optimized for point sources and therefore not neccesarily suitable for massive, nearby galaxies that are large and extended on the sky. One of the most impactful considerations is sky subtraction, which can over-subtract light in the outskirts of extended objects. This leads to total fluxes of massive galaxies being systematically underestimated ~\citep{Blanton2011,Fischer2017}. The Sloan Digital Sky Survey~\citep[ SDSS,][]{Abazajian2009} employs a drift-scan observing strategy, which helps control the systematic errors due to flat fielding and sky background.

Another important challenge to determining a galaxy's total flux is accounting for light from the galaxy that is below the noise limit of the observation. With a paramaterized fit, the light profiles of galaxies can be integrated analytically to infinity, but these results are sensitive to the chosen parameterization. Traditional methods, like those used to determine SDSS {\tt model} and {\tt cmodel} fluxes, which rely on rigid exponential and de Vaucouleurs profiles \citep{deVaucouleurs1949}, have been shown to severely underestimate the total flux of massive galaxies~\citep{Bernardi2013,Dsouza2015,Huang2018,Kravtsov2018}. Through re-analysis of the SDSS images, several studies suggest S\'ersic or two component bulge + disk models as more accurate tracers of the true light distribution of galaxies~\citep{Simard2011,Lackner2012,Kelvin2012,Meert2015,Bernardi2017a, Oh2017}. However, even when analyzing the same SDSS images, these studies disagree at the 10\% level mainly due to differences in the exact parameterization and how they estimate and subtract the sky background~\citep{Meert2015,Fischer2017}.

Massive galaxies are known to have a substantial fraction (10\% - 20\%) of their light in the low surface brightness outskirts, below roughly 25 mag arcsec$^{-2}$~\citep{vanDokkum2005,Tal2011,Duc2015,Iodice2016,Spavone2017}. This light is beyond the reach of large optical surveys, like SDSS. Therefore accurate stellar masses of massive galaxies require the correct paramaterization for the extrapolation of the light profile beyond the noise limits of the data. These choices can impact the calculated number density of massive galaxies by up to an order magnitude~\cite{Bernardi2013,Dsouza2015}

This is further complicated by the existence of extended diffuse light, often called the stellar halo or intra-halo light (IHL), which can extend out to the virial radius of a galaxy. This light is generally low surface-brightness, $\mu_g > 30\ \rm mag\ arcsec^{-2}$, and is the result of satellites which are disrupted by the tidal forces of the host galaxy and halo. Empirical models and hydrodynamical simulations suggest that this diffuse light represents a significant fraction ($\gtrsim 10\%$) of a galaxy's total stellar mass, generally becoming more important at higher masses ~\citep{Bullock2005,Conroy2007, Pillepich2018,Sanderson2018,Behroozi2019}. However, disagreement remains between predictions from hydrodynamical simulations and observations~\citep{Merritt2016,Monachesi2019,Merritt2020}. It is unclear how to, or even if one should, include this light as part of the total stellar mass of a galaxy.

%\note{Maybe give equations for m/l app or m/l tot}
Beyond the measurement of total flux, photometry across multiple photometric bands is required for use in spectral energy distribution (SED) fitting. A separate method is often used for this and then the results are ``normalized'' to the total flux measurment. One popular method is aperature photometry, which measure the flux within a fixed aperature. While these methods are able to produce consistent measurement over multiple photometic bands,
they implicitly ignore that galaxies have color gradients~\citep{Kormendy1989, Saglia2000,LaBarbera2005,Bakos2008, Tortora2010, Guo2012, Dominguezsanchez2019,Suess2020}. Given that these gradients in massive galaxies are generally negative (i.e. bluer colors at larger radii) the total color of a galaxy is likely to be bluer than that measured by aperture photometry. Other studies use paramaterized methods like the SDSS\texttt{model}~\citep{Ahn2014} or bulge + disk decompositions~\citep{Mendel2014}; however, these methods have additional issues, as discussed above.

%Common variants are the \textit{SExtractor}~\texttt{AUTO} method~\citep{Bertin1996} based on the Kron radius or the more advanced \texttt{LAMBDAR}~\citep{Wright2016} which aims to provide consistent photometry across many bands with varying depth and resolution. Since these methods measure the flux within a fixed aperture, generally $ 1\, r_{eff} -- 1.5\, r_{eff}$, 

In this study we test the commonly used methods for measuring the photometry of massive galaxies using an independent dataset from the Dragonfly Telephoto Array, Dragonfly for short~\citep{Abraham2014,Danieli2020}. Dragonfly currently consists of 48 telephoto lenses jointly aligned to image the same patch of sky in both the $g$ and $r$ bands. It operates as a refracting telescope with a 1m aperture and $f/0.4$ focal ratio. Dragonfly's design is optimized for low surface brightness imaging, routinely being able to image down to $\mu_g \gtrsim 30\rm \ mag\ arcsec^{-2}$ on 1 arcmin scales in the $g$ band~\citep{Merritt2016,Zhang2018,vanDokkum2019,Gilhuly2020}. SDSS remains the main dataset used to study massive galaxies in the local universe~\citep{Bernardi2017,Kravtsov2018} and Dragonfly offers a powerful complement to test the methods currently employed. It has superior large scale sky-subtraction due, in part, to the use of single CCDs that cover the entire $2.6^{\circ} \times 1.9^{\circ}$ FOV of each lens. Dragonfly's low surface brightness sensitivity allows the light profile to be measured to fainter limits, reducing the amount of extrapolation necessary thus minimizing the dependence on the choice of parameterization. 

We compare Dragonfly photometry to that of Galaxy Mass and Assembly survey \citep[GAMA,][]{Driver2011,Baldry2012}, and others. While GAMA is a spectroscopic survey at its heart, it also involved re-analyzing data from public imaging surveys, like SDSS, and performing own multi-wavelength imaging surveys. Our goal is to compare Dragonfly photometry to that published by GAMA~\citep{Kelvin2012,Wright2016} and other studies that re-analyze SDSS images~\citep[such as][]{Simard2011,Meert2015}. Specifically we focus on how these differences affect estimates of the total stellar mass and the measurement of the SMF.

The rest of the paper is organized as follows: In Section~\ref{sec:method} we describe and test our method for measuring the total flux of galaxies in the Dragonfly Wide Field Survey. Our galaxy sample is described in Section~\ref{sec:samp}, with initial results shown in Section~\ref{sec:results}. We compare our measurements to those of GAMA in Section~\ref{sec:res_GAMA} then investigate the effect of the differences on stellar mass estimates in section~\ref{sec:res_sm}. We compare Dragonfly measurements to other methods applied to SDSS images in Section~\ref{sec:meth_comp}. Our results are discussed in Section~\ref{sec:disc} and then summarized Section~\ref{sec:conc}.

\section{Measuring the photometry of galaxies in the DWFS}\label{sec:method}

\subsection{Data}
The main dataset we use in this study is the Dragonfly Wide Field Survey (DWFS), presented in ~\citet{Danieli2020}. This survey imaged $330\ \rm deg^2$ in well studied equatorial fields with superb low surface brightness sensitivity: the typical 1$\sigma$ depth is 31 mag arcsec$^{-2}$ on 10 arcmin scales. Our galaxy sample will be drawn from the GAMA database so we will focus on the part of the survey which overlaps with the GAMA equatorial fields~\citep{Baldry2018}. We use the final co-adds and refer the reader to \citet{Danieli2020} for details on the instrument, observations and data reduction. One particular detail of note is the sky subtraction procedure. It is performed in two stages, heavily masking all detected sources during the second stage, and fitting a 3rd order polynomial to the entire $1.8^{\circ} \times 1.2^{\circ}$ frame. This preserves any emission features on scales $\lesssim$ 0.6${^\circ}$ ensuring that the outskirts of galaxies are preserved for all the galaxies in our sample ($0.1< z < 0.2$, $0.5^{\prime \prime} \gtrsim r_{\rm eff} \gtrsim 8^{\prime \prime}$).

\begin{table}[h]
\centering
\begin{threeparttable}
\caption{The fields from the DWFS used in this study.}
\begin{tabular}{l|cc}
Field name & RA range (deg) & Dec range (deg) \\ \hline
G09\_130.5\_1 & 128.5 - 132.5 & -0.5 - 2.5 \\ 
G09\_136.5\_1$^*$ & 134.5 - 138.5 & -0.5 - 2.5 \\ 
G09\_139.5\_1$^*$ & 137.5 - 141.5 & -0.5 - 2.5 \\ 
G09\_139.5\_m1$^*$ & 137.5 - 141.5 & -2.5 - 0.5 \\ 
G12\_175.5\_1$^\dag$ & 173.5 - 177.5 & -0.5 - 2.5 \\ 
G12\_175.5\_m1$^\dag$ & 173.5 - 177.5 & -2.5 - 0.5 \\ 
G12\_178.5\_m1$^\dag$ & 176.5 - 180.5 & -2.5 - 0.5 \\ 
G15\_213\_1 & 211.0 - 215.0 & -0.5 - 2.5 \\ 
G15\_222\_1$^\ddag$ & 220.0 - 224.0 & -0.5 - 2.5 \\ 
G15\_222\_m1$^\ddag$ & 220.0 - 224.0 & -2.5 - 0.5 \\
\hline
\end{tabular}
\begin{tablenotes}
\footnotesize
\item $^*,^\dag,^\ddag$ indicate fields with overlap.
\end{tablenotes}
\label{tab:fields}
\end{threeparttable}
\end{table}

The specific fields we will be using are shown in Table~\ref{tab:fields}. Each field is $4^\circ \times 3^\circ$, which is larger then the FOV of a single DF lens due to the dithering pattern adopted for the DWFS~\citep{Danieli2020}. These fields overlap with the G09, G12 and G15 GAMA fields. The ten DWFS fields represent 104 $\rm deg^2$, with overlap between several of fields. These overlapping regions, totalling roughly 16 $\rm deg^2$, will be useful later to test and validate our measurements. The DWFS images have a pixel scale of $2.5\rm\ ^{\prime \prime}/\, pixel$ and typical full width at half maximum (FWHM) of the point spread function (PSF) is $5^{\prime \prime}$

\begin{figure*}
    \centering
    \includegraphics[width = 0.9\textwidth]{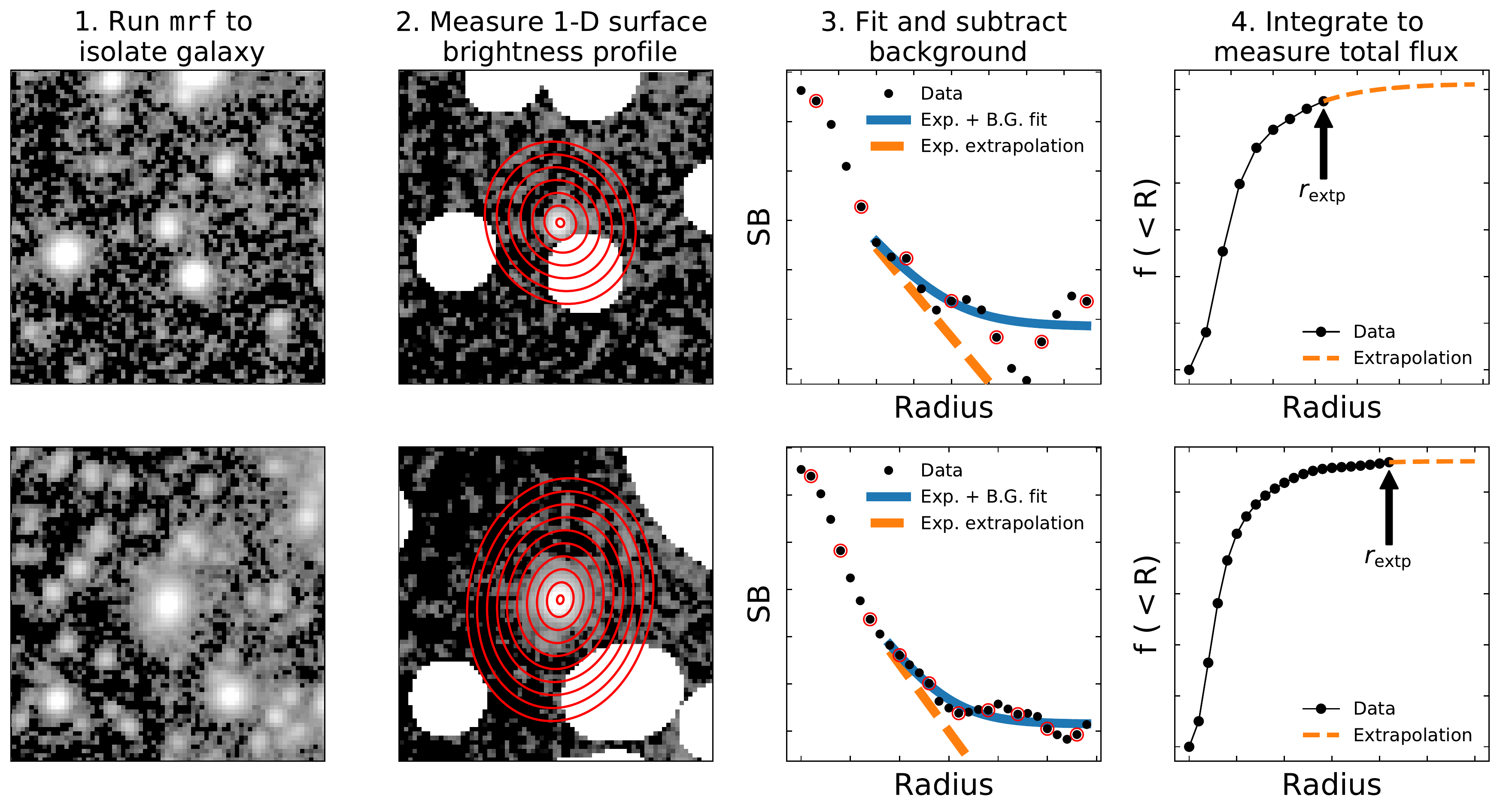}
    \caption{Illustration of the four major steps involved in our method of measuring the total flux of galaxies in the DWFS. For full details see Section~\ref{sec:method} of the text. We show two example galaxies in the two rows. \textbf{1st Column:} The raw Dragonfly $r$ band images are shown. \textbf{2nd:} The Dragonfly images after the \texttt{mrf} procedure. The red lines show the isophotes along which the surface brightness profile is measured. \textbf{3rd:} The measured surface-brightness profile is shown. The blue solid and orange dashed lines show the exponential + background fit and just the exponential portion respectively. \textbf{4th:} The measured curve of growth is shown. The orange dashed line shows the extrapolation calculated using the exponential fit. $r_{\rm extp}$, the radius beyond which we extrapolate the profile is also shown.}
    \label{fig:method}
\end{figure*}

%\note{Move this paragraph?}
We develop a method that allows for a non-parametric measurement of the photometry of galaxies. The method is summarized in Figure~\ref{fig:method} and will be described in detail below. Given the limitations of the Dragonfly data, mainly the poor spatial resolution, we make quality cuts at certain steps during the method, where the photometry of certain galaxies cannot be measured accurately. Thus, the result is not a complete sample of all galaxies in the field. As we show below, we verify that these cuts do not introduce significant biases.

\subsection{Using \texttt{mrf} to isolate galaxies}

The first step in our method is isolating the galaxies from other emission using the Multi-resolution filtering (\texttt{mrf}) algorithm\footnote{\url{https://github.com/AstroJacobLi/mrf} }. This algorithm, developed by \citet{vanDokkum2019a}, is designed to isolate low surface brightness features in low-resolution data, like Dragonfly, by using an independent, higher resolution image to remove compact, high surface brightness emission. A kernel is derived to match the PSFs of two datasets with different spatial resolution. The high-resolution data is then degraded to the resolution of Dragonfly and used to remove the emission due to compact sources.

For the high resolution data we use images from the The Dark Energy Camera Legacy Survey (DECaLS) \citep{Dey2019} data-release 8 \footnote{Downloaded from \url{http://legacysurvey.org/dr8/}}. The \textit{DECaLS} images act as the high-resolution data (pixel scale = $0.262\ ^{\prime \prime }/\rm pix$) to run \texttt{mrf} on our DWFS images (pixel scale = $2.5\ ^{\prime \prime}/\rm pix$). We use $ 0.7^{\circ} \times 0.7^{\circ}$ cutouts of the DWFS and \textit{DECaLS} data centered on each galaxy. This is a large enough area, containing enough unsaturated stars for the \texttt{mrf} algorithm to consistently derive an accurate kernel between the two images. Regions where very bright objects have been removed are additionally masked. We use the masked results of this \texttt{mrf} procedure in all the following analysis.
%% Best way to supply YAML file.

\subsection{Measuring the 1-D surface brightness profile}

Once we have run \texttt{mrf} on the galaxy cutout we then measure the 1-D surface brightness profile. This is done using the python package \texttt{photutils}\footnote{https://github.com/astropy/photutils} \citep{photutils}. In particular we measure the surface brightness profile using the \texttt{ELLIPSE} method based on the algorithm developed in \citet{Jedrzejewski1987}. This is performed in a non-iterative mode using the sky positions, position angles and axis ratio for each galaxy determined in the GAMA S\`{e}rsic photometry of the SDSS images~\citep{Kelvin2012}. These quantities are fixed at all radii. To ensure the axis ratio is applicable for DWFS observations, where the resolution is roughly 10 times lower, we ``convolve'' the intrinsic axis ratio following \citet{Suess2019}:

\begin{equation}
    b/a _{\rm obs.} = \sqrt{ \frac{(b/a_{\rm intr}\, r_{\rm eff} )^2 + r_{psf}^2 }{{r_{\rm eff}}^2 + r_{psf}^2} }
\end{equation}

Here, $b/a_{\rm intr}$ is the intrinsic axis ratio, and $r_{psf}^2$ is the PSF HWHM, which we assume to be $2.5^{\prime \prime}$ for the DWFS, and $r_{\rm eff}$ is the half-light radius measured from the S\'ersic fits performed on the SDSS image by~\citet{Kelvin2012}. We measure the surface brightness profile in 2.5 arcsec (corresponding to 1 Dragonfly pixel) steps from the center of the galaxy out to $20\ r_{\rm eff}$. During this procedure we discard galaxies where the algorithm fails to converge, which happens for roughly 30\% of galaxies. This generally occurs because a significant fraction of pixels are masked near the center of the galaxy. 

\subsection{Background subtraction}
We use the 1-D surface brightness profile to calculate the total flux of each galaxy. First a constant sky background is subtracted from the profile. This is done by fitting an exponential plus a constant background to the outskirts of the galaxy profile as follows,

\begin{equation}
    \label{eqn:bg_model}
    I(r)\ =\ \alpha\, e^{ -\beta*r} + c
\end{equation}{}

Here $\alpha,\beta\ {\rm and}\ c$ are the free parameters to be fit. We only fit regions of the galaxy that have surface brightness  $\geq 28.5\ \rm mag\ arcsec^{-2}$ in either band. In the $r$ band this typically occurs at radii larger then $ 7\times r_{\rm eff}$. For a small fraction of galaxies, this fit does not converge, and these galaxies are discarded.  The constant background ($c$) is subtracted from the entire surface brightness profile, which is then used in the next step to calculate the total flux. We test this method below, in Section~\ref{sec:vtest}, with the injection of artificial galaxies and show that this method successfully subtracts the background regardless of the galaxy profile.

\subsection{Measurement of total flux}

Next, we find where the background-subtracted surface brightness profile first drops below a signal to noise ratio (SNR) of 2. We call that radius $r_{\rm extp.}$. We integrate up to this point to calculate the observed flux, $S_{\rm obs}$, as shown below:
\begin{equation}
    S_{\rm obs.} = 2\pi(b/a _{\rm obs}) \int_0^{r_{\rm extp} } r\, f(r)\, dr
\end{equation}
Here, $f(r)$ is the background subtracted surface brightness profile, and the integral is performed using a simple trapezoidal rule. Although we aim to be non-parametric, we still need to extrapolate the profile to account for any light beyond $r_{\rm extp}$. To accomplish this we use the exponential part of the fit described in Eqn. \ref{eqn:bg_model}, and integrate from $r_{\rm extp}$ to infinity. This flux we call the extrapolated flux, and it is calculated as:

 \begin{equation}\label{eqn:extp_int}
 \begin{split}
S_{\rm extp} &= 2 \pi (b/a _{\rm obs}) \int_{r_{\rm extp} }^{\infty} r\, \alpha\, \exp( -\beta\, r) dr\\ 
&= 2 \pi\, (b/a _{\rm obs})\, \alpha\, e^{ -\beta\, r_{\rm extp} } \left( \frac{ 1 + \beta\, r_{\rm extp} } {\beta^2} \right)
 \end{split}
\end{equation}

Here, $\alpha$ and $\beta$ are the best fit values from the model of the galaxy profile, as shown in Equation~\ref{eqn:bg_model}. 

We have made a choice to use an exponential extrapolation, but it remains our largest systematic uncertainty.  We show below in Section~\ref{sec:vtest} through comparing results from overlapping regions and injection-recovery tests that it produces accurate and reliable results for S\'ersic profiles. However, this does not necessarily reflect reality. This issue is further discussed in Section~\ref{sec:disc}. We also tested a power-law extrapolation. In this case we restrict the power law slope to $< -3$, to ensure the integral converges. This achieved similar results to the exponential extrapolation. We also tried a S\'ersic like profile with an additional shape parameter, $n$. However, the fit failed due to the data not being able to constrain the $n$ parameter consistently. 

Finally the data are brought to the same photometric system as the SDSS photometry. Even though the $g$ and $r$ filters used on Dragonfly are very similar to those of SDSS,  a small correction needs to be applied. We derive corrections for the $g$ and $r$ bands in Appendix~\ref{sec:filt_conv} by comparing the photometry of standard stars in SDSS, Dragonfly and Gaia~\citep{Gaia2018}. For all of the galaxies we convert the Dragonfly magnitudes to the SDSS filter system based on the observed Dragonfly colors. The median corrections are $\Delta_g = 0.06$ mag and $\Delta_r =  0.03$ mag for the $g$ and $r$ bands, respectively, where $m_{\rm SDSS} = m_{\rm DF} + \Delta_x$. The Dragonfly colors are calculated using the total flux measurements once they have been adjusted to match the SDSS filter system.

\subsection{Validation tests}
\label{sec:vtest}
We perform two separate tests of our method. The first is an injection-recovery test. We generate 1500 single component S\'ersic models~\citep{Sersic1963} using \texttt{galsim}\footnote{https://github.com/GalSim-developers/GalSim} with sizes, total magnitudes, S\'ersic indicies and axis ratios drawn from a distribution similar to the observed galaxies. We inject the models into the DWFS frames and perform the entire analysis pipeline, described in section Sec.~\ref{sec:samp}, on the simulated galaxies, including all the quality cuts and the \texttt{mrf} procedure. After all the quality cuts, including the bright neighbour cut described below in Sec.~\ref{sec:samp}, the photometry is measured for roughly $60 \%$ of the simulated galaxies. 
 
\begin{figure}
    \centering
    \includegraphics[width = 0.9\columnwidth]{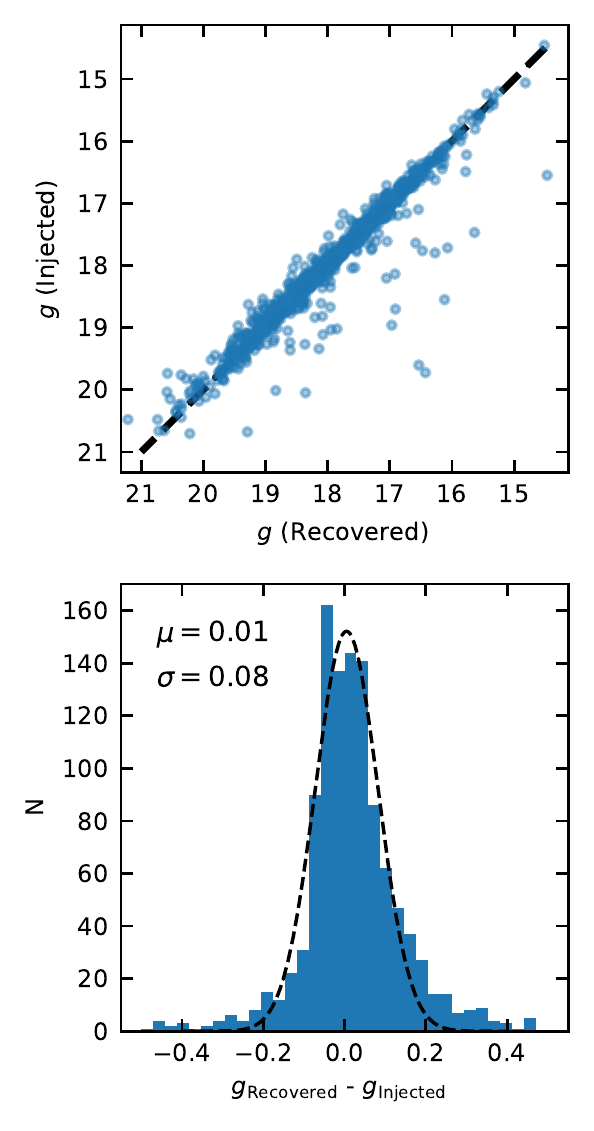}
    \caption{\textbf{Top:} Comparison of injected and recovered single component S\'ersic models using our pipeline. The black dashed line shows the one-to-one relation. We find our method accurately recovers the input magnitude well for all galaxies except a few outliers where the recovered magnitude is much greater than the input. \textbf{Bottom:} The distribution of differences between the injected and recovered magnitudes. The black dashed line shows a Gaussian fit to the distribution.}
    \label{fig:inj_rev}
\end{figure}

Figure~\ref{fig:inj_rev} displays the results of these tests, focusing on the $g$ band. We find that our non-parametric method works well in general. The mean of the $m_{\rm recovered} - m_{\rm injected}$ distribution is near zero, $\mu = 0.005 \pm 0.004$. suggesting there is no systemic bias, and a scatter of 0.08 mag. We note that the distribution appears slightly skewed to positive magnitude differences. There is also a small fraction of outliers ($\sim 3\%$) for which the recovered magnitude is much brighter then the injected magnitude. This seems to be caused by nearby bright objects that are just outside the $10\, r_{\rm eff}$ cut-off or not present in the SDSS catalog. When we increase the size of this cut-off to $20\, r_{\rm eff}$, the fraction of outliers decreases, but the total number of galaxies in our sample drastically drops. Therefore we decided to keep the cut-off at $10\, r_{\rm eff}$ and accept that there will be some outliers.

\begin{figure}
    \centering
    \includegraphics[width = \columnwidth]{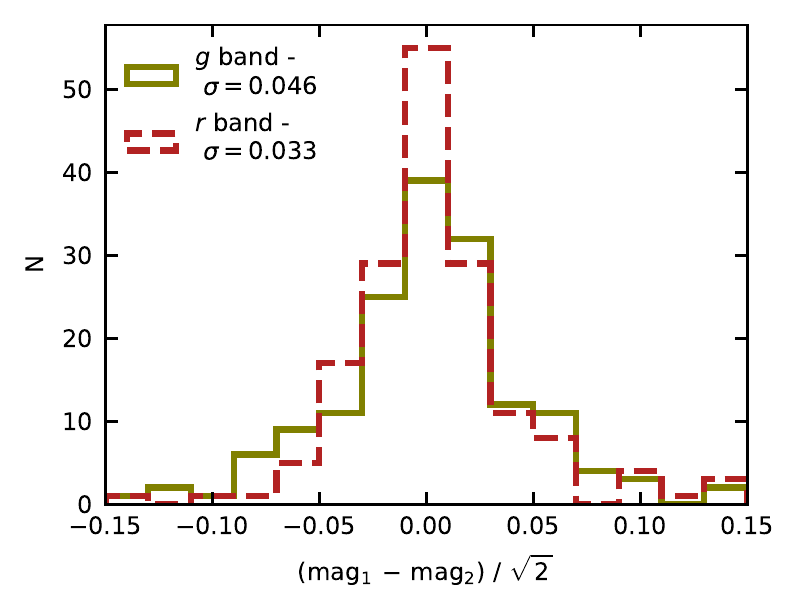}
    \caption{Comparison of independent measurements of the same galaxies which are present in two different fields. We divide the magnitude difference by $\sqrt{2}$ as we are using the differences to probe the underlying uncertainty of a given measurement. }
    \label{fig:dups}
\end{figure}

The second test we performed is to compare the photometry of galaxies which lie in the region of overlap between multiple survey fields. These galaxies have had their photometry independently measured and represent a good test of the reliability and uncertainty of our method. There are 169 galaxies which have their photometry successfully measured in multiple fields. We note that the overlap is naturally at the edges of the DWFS fields where the noise is higher, therefore this represents a conservative test. The distribution of magnitude differences between the independent measurements of the same galaxy is shown in Figure~\ref{fig:dups}. We show the magnitude differences divided by $\sqrt{2}$, that is, the typical $1\sigma$ uncertainty for each galaxy, assuming a Gaussian error distribution.

This distribution is centered at 0 with a width of $\sigma = 0.046$ mag for the $g$ band and $0.033$ mag for the $r$ band, calculated as the bi-weight scale. Similar to the distribution of recovered magnitudes above, the distribution of the magnitude differences is well approximated by a Gaussian near the center, however there are outliers. Specifically 8\% (4\%) of this sample has $|\Delta\ \rm mag\, |/ \sqrt{2} > 0.15$ in the $g$ ($r$) band, which greatly exceeds the expectation of a Gaussian distribution. 

We note that the uncertainty implied by comparing measurements of the same galaxy is significantly smaller than that implied by the injection-recovery tests. In a sense they are measuring two different things. Comparing independent measurements takes the uncertainties in the Dragonfly data, reduction and calibration into account. On the other hand, the injection-recovery test probes the systematic uncertainty caused by, among other aspects of our methods, our choice of an exponential extrapolation.

In short the error in our photometry measurements is still uncertain and depends on the true surface-brightness profile in the outskirts of galaxies.  Specifically the systematic error caused by our choice of an exponential extrapolation and how well it matches realistic galaxy profiles. We plan to investigate this in future works using Dragonfly data by stacking.

\section{Galaxy Sample}
\label{sec:samp}
The construction of the galaxy sample relies on the GAMA survey, \citep[specifically DR3, ][]{Baldry2018}. We select galaxies within the DWFS footprint with $\log M_*/M_\odot> 10.75$ and $0.1 < z < 0.2$, along with the additional quality cuts: $\texttt{SpecAll.nQ} > 2$ and $\texttt{StellarMasses.nbands} > 3$ as suggested on the data access website\footnote{http://www.gama-survey.org/dr3/schema/}. This mass-selected galaxy sample is $> 99\%$ complete~\citep{Taylor2011}. The parent sample contains 3979 galaxies over the ten DWFS fields used.

Next we remove galaxies that have nearby bright objects. This step is done to avoid confusion of the galaxy's light with light from nearby objects. We use the \texttt{PhotoObjAll} table from the SDSS DR14 database\footnote{\url{http://skyserver.sdss.org/dr14/}} to search for sources nearby the galaxies in question. If there is another source that is at least 0.5 times as bright in either the $g$ or the $r$ band within $10 \times r_{\rm eff}$ in either the $g$ or $r$ band \citep[as measured by the S\'ersic fits in ][]{Kelvin2012} then the galaxy is discarded. This selection is done to remove galaxies in close pairs or nearby bright stars which could contaminate the photometry of the galaxy. This cut removes about $40 \%$ of the galaxies from the parent sample. 

We obtain photometry for the remaining galaxies. During this process an additional $\sim 30 \%$ of the parent sample is discarded during the process of measuring the isophotes. This is often due to there being too many masked pixels near the centre of the galaxy. Finally another small fraction, $< 2 \%$, of the galaxies in the parent sample is discarded because the exponential + background fit to the outskirts of the 1-D spectrum does not converge. As a final cut, we do not include anything where the extrapolated fraction of the total flux, $ f_{\rm extp} =  S_{\rm extp}/  (S_{\rm extp} + S_{\rm obs.})$ exceeds 10\%. This is done to remove galaxies with spurious background fits or that are low signal-to-noise in the Dragonfly data. Overall, this cut removes a small fraction of galaxies ($< 1\%$) from the parent sample of galaxies. For the remainder of this paper, we use the analysis galaxy sample for which the photometry is fully measured as described above, containing 1188 galaxies, about 30\% of the parent sample.

\begin{figure*}
    \centering
    \includegraphics[width = \textwidth]{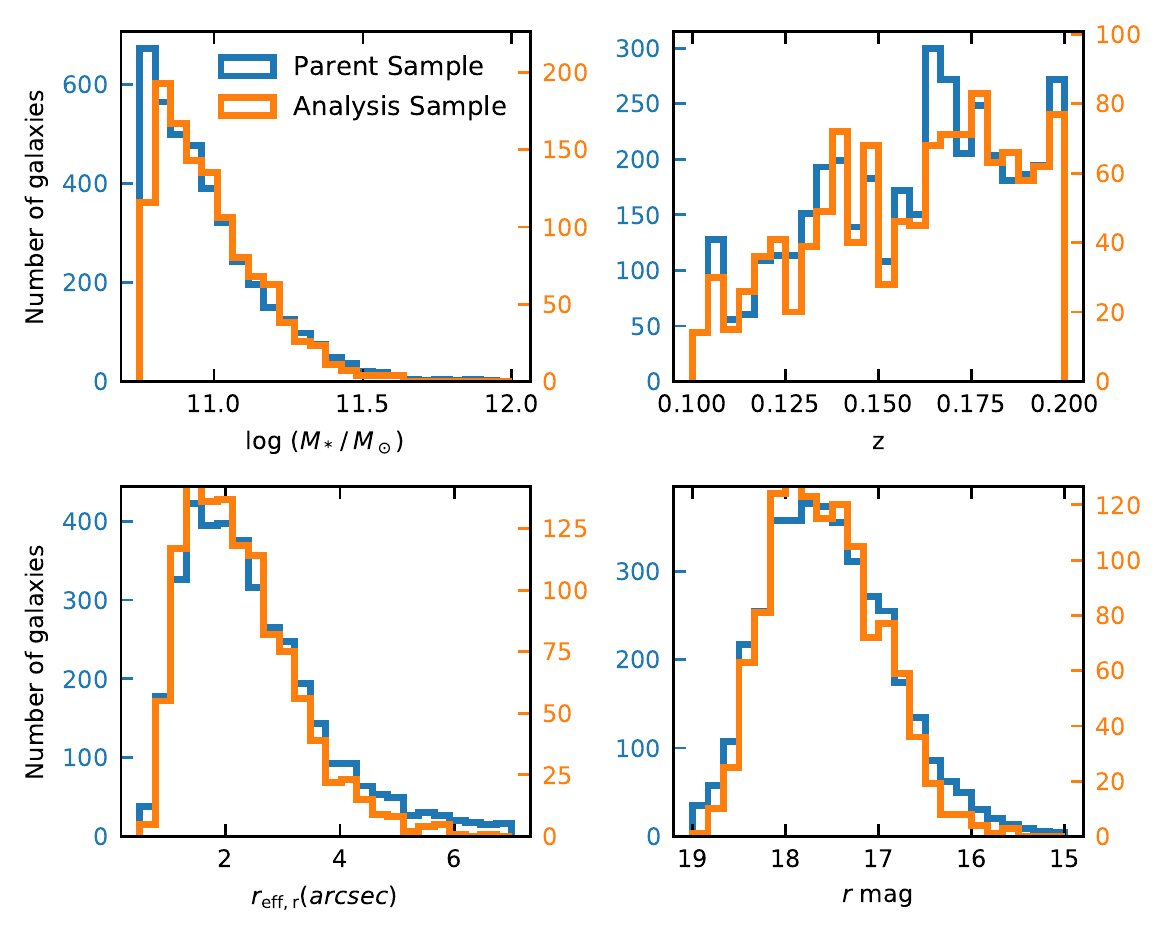}
    \caption{Distributions of galaxy properties for the parent sample of galaxies (Blue), drawn directly from the GAMA DR3 database, and our analysis sample (Orange) which has passed all of the quality cuts. The effective radius and magnitude measurements come from the 2-D S\'ersic fits performed by~\citet{Kelvin2012} on SDSS data. We find the distributions of the two samples are similar, suggesting that the analysis sample is representative of the parent sample.}
    \label{fig:samples_hist}
\end{figure*}

Figure~\ref{fig:samples_hist} displays the distribution of galaxy proprieties for both the parent and analysis galaxy samples. The analysis sample contains galaxies that have passed all the quality cuts and the Dragonfly photometry has been successfully measured in both bands. While the analysis sample only contains $\sim 30\%$ of the parent sample, the distributions of galaxy properties are similar. This implies that we have not biased the sample significantly while performing the quality cuts described in the method above, and it is representative of the parent sample.

\begin{figure*}
    \centering
    \includegraphics[width = 0.85\textwidth]{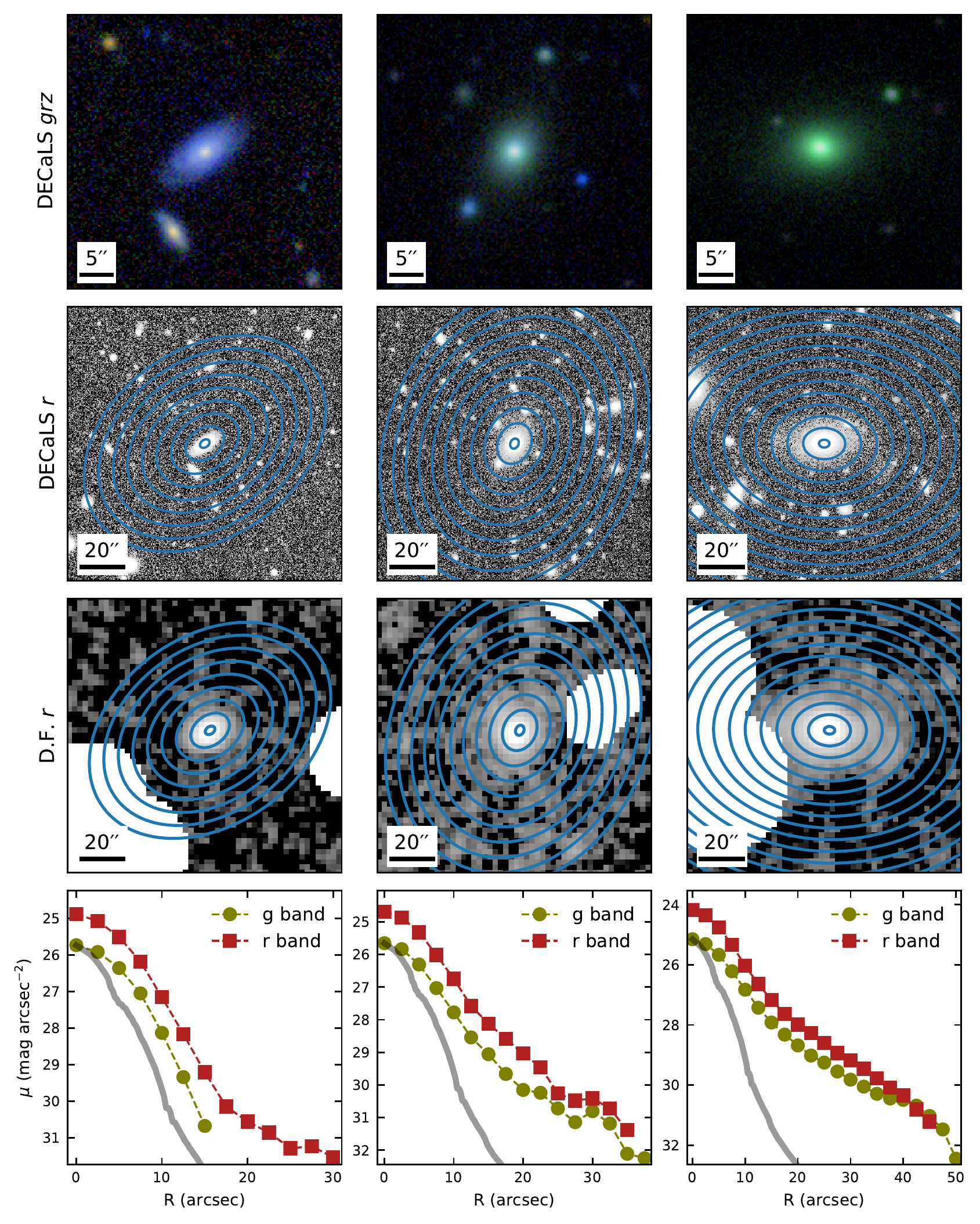}
    \caption{Imaging of example galaxies from \textit{DECaLS} and Dragonfly. All 3 galaxies have $\log M_* / M_\odot \sim 11$ and $z = 0.19,0.15,0.12$ from left to right. \textbf{Top}: False color image based on \textit{DECaLS} $g$, $r$ and $z$ imaging. \textbf{Second Row:} \textit{DECaLS} $r$ band image logarithmic scaling to highlight the ourskirts of the galaxy. White displays 23 mag arcsec$^{-2}$ and black shows 32 mag arcsec$^{-2}$ \textbf{Third Row:} Dragonfly $r$ band images after performing the \texttt{mrf} procedure, with scaling to match the surface brightness limits of \textit{DECaLS} image above. Blue contours are placed every $7.5$ arcsec (3 DF pixels) and are matched between the two images. \textbf{Bottom}: Background subtracted surface brightness profiled measured from the Dragonfly data. The grey line shows an example of the Dragonfly PSF.}
    \label{fig:sbp}
\end{figure*}

\section{Photometry measurements from Dragonfly}
\label{sec:results}
In Figure~\ref{fig:sbp} we show some example galaxies in both high-resolution  and Dragonfly data. We show $grz$ false color images along with logarithmicaly stretched deep $r$ band images from DECaLS. The false color image is zoomed to roughly 1 arcmin per side while the r-band images is 2 arcmin per side. To compare directly, we plot the Dragonfly r-band images using a logarithmicaly strecth matching the surface-brightness limits of the DECaLS image. The isophotes are matched across the two images.

Comparing the DECaLS and Dragonfly images illustrates two important points. First, Dragonfly has a much poorer spatial resolution compared to most modern optical surveys. Therefore, we are not able to spatially resolve the centers of galaxies on scales $\leq 5^{ \prime \prime}$. However, the strength of Dragonfly is in studying the low surface-brightness outskirts of these galaxies. Even visually, one can see that the galaxies in the Dragonfly images are more extended, as a result of the superior low surface-brightness performance compared to the high-resolution data.

The final row shows the measured $g$ and $r$ band surface brightness profiles measured in the DWFS along with an example of the g band PSF. In detail, the PSF varies between the bands, different fields and different observing nights (see Liu et al. in prep), but it remains qualitatively similar. Comparing the galaxy profile to the PSF further demonstrates that many of the galaxies in our sample are barely resolved. However, this should not affect our measurements of the total flux. One of the advantages of Dragonfly is that it has an extremely well controlled PSF with very little scattered light at large radii, therefore we are able to recover the total flux of galaxies even if they are not well resolved.

\begin{figure}
    \centering
    \includegraphics[width = \columnwidth]{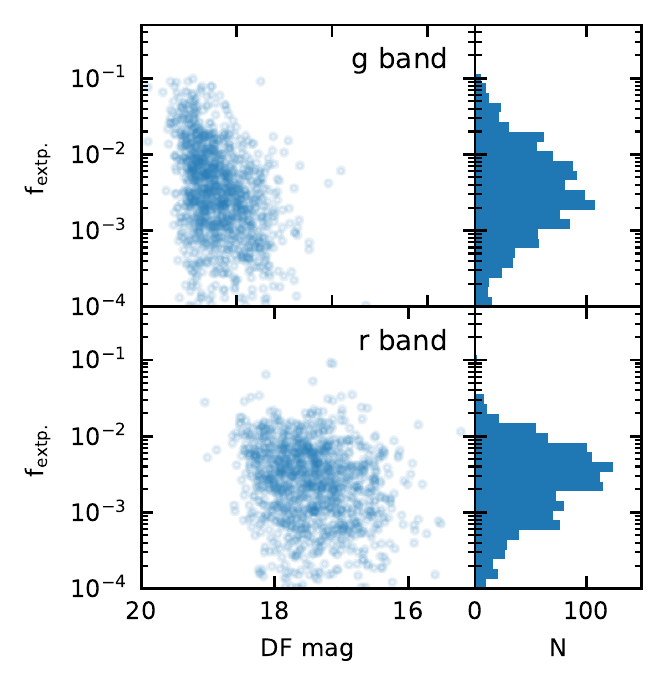}
    \caption{The fraction of total flux contained in the exponential extrapolation, $f_{\rm extp.}$, as a function of the DWFS measured magnitude for both the $g$ and $r$ band. In both bands $f_{\rm extp.}$ approximately follows a log-normal distribution with a mean of 10$^{-2.5}$ and width of 0.6 dex. }
    \label{fig:f_extp}
\end{figure}

Another way to asses our pipeline is to investigate the fraction of the total flux contained in the exponential extrapolation, $f_{\rm extp}$. Ideally this will be very small, so that the details of the extrapolation do not affect the total flux measurement significantly. In Figure~\ref{fig:f_extp} we show both the $g$ and $r$ extrapolated fraction as a function of the observed magnitude.

In both the $g$ and $r$ bands, $f_{\rm extp}$ has a log-normal distribution with a mean of roughly $10^{-2.5}$ and standard deviation of 0.6 dex. Interestingly, there does not appear to be a strong correlation with the observed magnitude of the galaxy. $f_{\rm extp}$ is generally below 1\% implying that we are measuring more than 99\% of the total flux of a galaxy with Dragonfly. This result is somewhat dependent on the form of extrapolation but it is nonetheless encouraging. 
We make the catalog of photometry measurements for the final analysis sample publicly available here\footnote{\url{https://tbmiller-astro.github.io/data/}}

\section{Comparing Dragonfly photometry to GAMA}
\label{sec:res_GAMA}

In this section we will be comparing the photometry observed in the DWFS to that measured by GAMA DR3. We will be comparing the DWFS total flux measurements to GAMA measurements that use S\'ersic photometry, truncated at $10\ r_{\rm eff}$. We also compare the Dragonfly color, measured using the total fluxes, to the \texttt{AUTO} color~\citep{Kelvin2012,Driver2016}. While the database contains a more sophisticated \texttt{LAMBDAR} photometric measurements that performs better in the UV and IR, where the resolution differ greatly from the optical, for the $g$ and $r$ colors it produces nearly identical results to the \texttt{AUTO} measurements~\citep{Wright2016}. We have chosen to focus on these methods of measuring the total flux and color as they form the basis of the stellar mass measurements used to calculate the SMF in \citet{Baldry2012} and \citet{Wright2017}.

\subsection{Comparing the total flux to S\'ersic Photometry}
\label{sec:res_sersic}
The first comparison we will be making in this paper is the total flux measured by Dragonfly to the S\'ersic fits performed by \citet{Kelvin2012}. The single component S\'ersic models were found by running \texttt{GALFIT} on SDSS imaging data. The $r$ band S\'ersic model, truncated at $10\ r_{e}$, is used by \citet{Baldry2012} and \citet{Wright2017} as the total flux normalization of stellar masses of galaxies measured from SED fitting. Here we compare to the truncated S\'ersic magnitudes in the $g$ and $r$ band. We will refer to these magnitudes as $g_{\rm GAMA}$ and $r_{\rm GAMA}$

\begin{figure*}
    \centering
        \includegraphics[width = 0.98\textwidth]{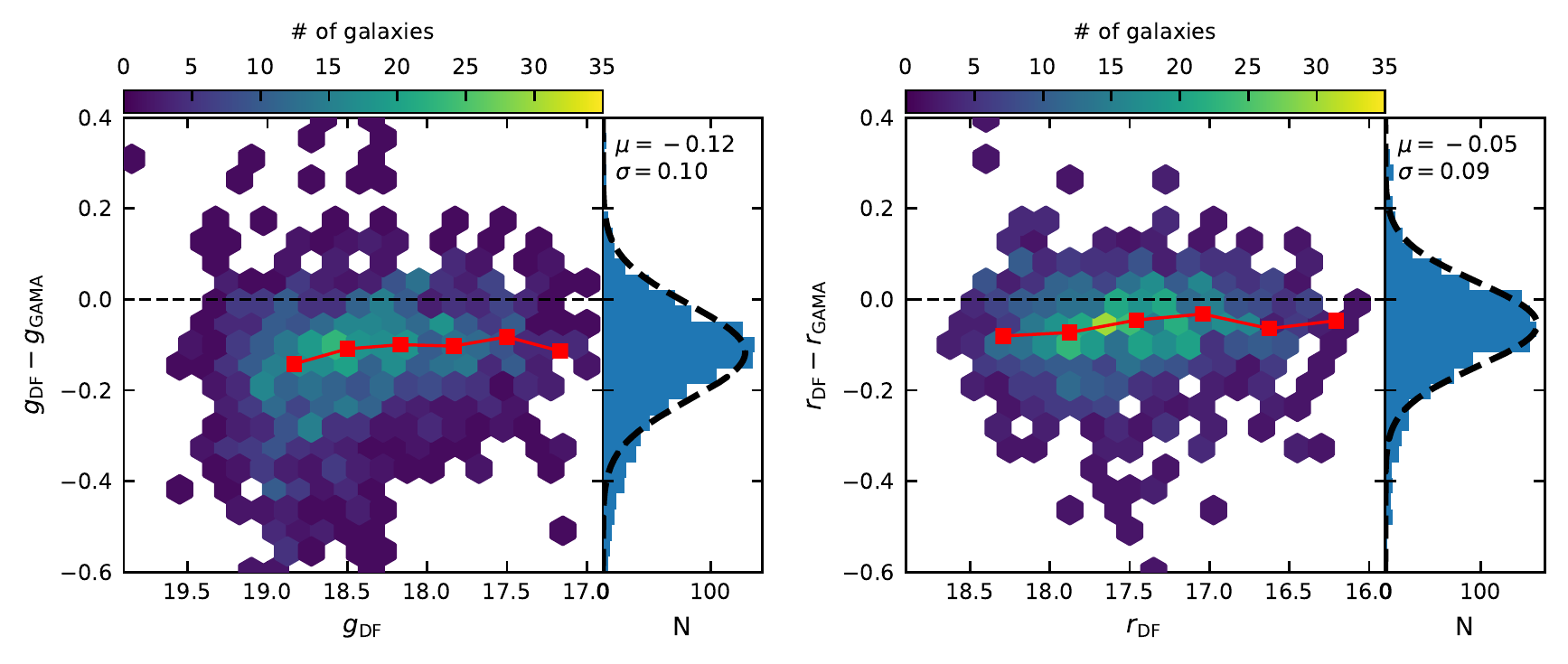}
    \caption {Comparison of Dragonfly to S\'ersic photometry performed by~\citet{Kelvin2012} as part of the GAMA survey. We show the distribution of the magnitude differences between our Dragonfly and GAMA magnitudes as a function the Dragonfly magnitude. The red squares show the median magnitude difference as a function of Dragonfly magnitude. Both bands follow a roughly Gaussian distribution. The black dotted line in each histogram panel shows the result of a Gaussian fit to the distribution with parameters displayed in each panel. On average galaxies are brighter in Dragonfly compared to the GAMA S\'ersic measurements}
    \label{fig:dmag}
\end{figure*}

In Figure~\ref{fig:dmag} we compare DWFS to the GAMA S\'ersic photometry for both the $g$ and $r$ bands. In both the $g$ and $r$ band, we find that on average, galaxies are brighter in Dragonfly, i.e. have negative $m_{\rm DF} - m_{\rm GAMA}$. In the $r$ band,  we find there is little dependence of $r_{\rm DF} - r_{\rm GAMA}$ on observed magnitude. For the $g$ band the difference increases for galaxies with $g_{\rm DF} > 18.5$. For bright galaxies, there is no dependence on $g_{DF}$ but for fainter galaxies, $g_{\rm DF} - g_{\rm GAMA}$ continues to decrease.

\begin{figure}
    \centering
    \includegraphics[width = 0.99 \columnwidth]{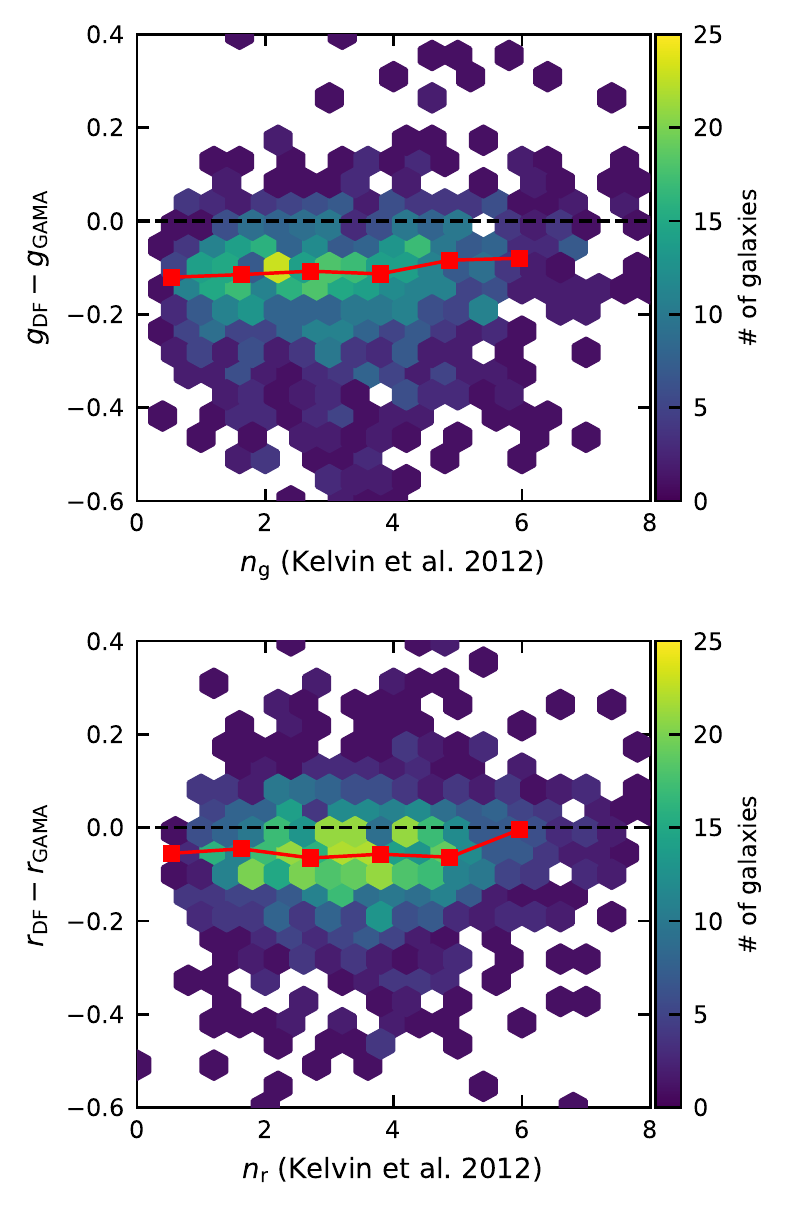}
    \caption{The difference between Dragonfly and GAMA S\'ersic photometetry as a function of S\'ersic index. The S\'ersic index is calucalated in \citet{Kelvin2012} using SDSS data.The red squares show the median magnitude difference as a function of S\'ersic index. There is no significant trend with S\'ersic index}
    \label{fig:dmag_n}
\end{figure}

The overall distributions of $g_{\rm DF} - g_{\rm GAMA}$ and $r_{\rm DF} - r_{\rm GAMA}$ are also shown in Figure~\ref{fig:dmag}. Each distribution is roughly Gaussian with a mean of -0.12 in the $g$ band and $-0.05$ in the $r$ band. The width of both distributions are similar with $\sigma = 0.10$ and $\sigma = 0.09$ for the $g$ and $r$ band respectively.

Since we are using an exponential extrapolation (see Sec.~\ref{sec:method}) one might expect there to be a systematic bias as a function of the S\'ersic index, such that high $n$ galaxies have their total flux underestimated. On the other hand, possible truncations in the light profile \citet{Pohlen2000,Trujillo2005} could work in the opposite direction. To investigate whether there is a correlation with a galaxy's structure in Figure~\ref{fig:dmag_n} we show how the difference between dragonfly and GAMA S\'ersic photometry depends on the measured S\'ersic index of a galaxy. We do not find that there is a significant trend, suggesting our method is robust against such a systematic effect. Although our exponential extrapolation doesn't match the true profile of high $n$ galaxies, since $f_{\rm extp}$ is $< 10^{-2}$ for many of our galaxies, this potential bias is minimized.

\subsection{Comparing $g-r$ colors to aperture photometry}
\label{sec:res_auto}

\begin{figure}
    \centering
    \includegraphics[width = \columnwidth]{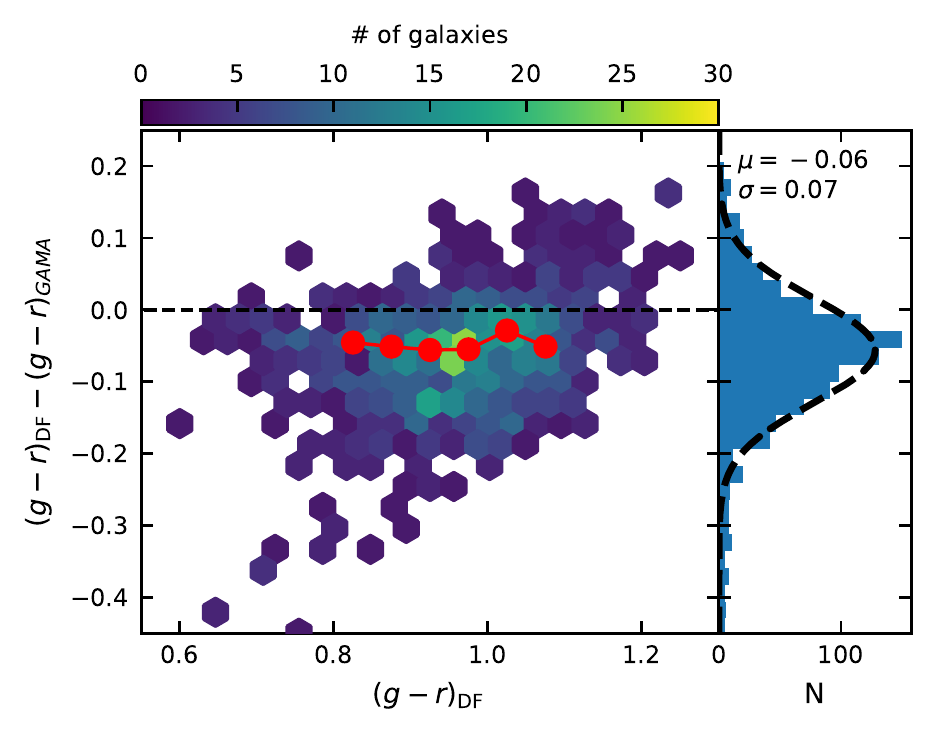}
    \caption{Comparison between Dragonfly measured $g-r$ colors to GAMA colors measured by~\citet{Driver2016} using the aperature matched \texttt{AUTO} method on SDSS data. The red line displays the running median color difference as a function of $(g-r)_{\rm DF}$. The black dashed line in right panel shows the result of a Gaussian fit to the distribution with parameters: $\mu = -0.05$ and $\sigma = 0.08$. In general, Dragonfly measures bluer colors compared to the aperture matched technique.}
    \label{fig:dcol}
\end{figure}

Here, we will be comparing Dragonfly measured $(g-r)$ color to aperture matched photometry. The Dragonfly colors are calculated using the total flux measurements. We will be comparing to the GAMA \textit{SExtractor} \texttt{AUTO} photometry performed on SDSS images~\citep{Bertin1996, Driver2016}. The flux is measured across the multiple bands within the Kron radius of the $r$-band image, typically $ 1\ r_{eff, r}\ \text{--}\ 1.5\ r_{eff, r}$~\citep{Kron1980}. This color will be referred to as $(g-r)_{\rm GAMA}$.

%This is used to obtain consistent photometry across multiple bands when measuring colors or to use for SED fitting. One band is chosen as the reference bands and defines the aperture for each source, measurements from the other bands are then taken from the same fixed aperture to ensure consistent photometry. Specifically we will be comparing to the GAMA \texttt{AUTO} photometry performed on SDSS images where $r$ is used as the primary detection band~\citep{Driver2016}.

The difference between GAMA and Dragonfly measured  $g-r$ colors is shown in Figure~\ref{fig:dcol}. We find there is little dependence of $(g-r)_{\rm DF} - (g-r)_{\rm GAMA}$ on the observed color. While there may appear to be a correlation, especially at the extremes of the distribution, it is important to remember that $(g-r)_{\rm DF}$ is plotted on both the $x$ and $y$ axis. Therefore this apparent correlation is likely caused by outliers in the $(g-r)_{\rm DF}$ distribution and does not reflect an inherent relationship. Also shown is the distribution of $(g-r)_{\rm DF} - (g-r)_{\rm GAMA}$ for all galaxies in the analysis sample. The distribution appears relatively Gaussian with a mean of $-0.06$ mag, consistent with the difference between $g_{\rm DF} - g_{\rm Sersic}$ and $r_{\rm DF} - r_{\rm Sersic}$ shown in Figure~\ref{fig:dmag}.

Interestingly the width of the distribution of $(g-r)_{\rm DF} - (g-r)_{\rm GAMA}$ is smaller than that of $g_{\rm DF} - g_{\rm GAMA}$ or $r_{\rm DF} - r_{\rm GAMA}$ individually. While not explicitly measuring the same thing, this implies there is some correlation between $g_{\rm DF} - g_{\rm GAMA, AUTO}$ and $ r_{\rm DF} - r_{\rm GAMA, AUTO}$. Indeed, we calculate the Pearson correlation coefficient between  $g_{\rm DF} - g_{\rm GAMA, AUTO}$ and $ r_{\rm DF} - r_{\rm GAMA, AUTO}$ to be 0.4, suggesting a moderate correlation. 

To gain insight into what is causing the difference between the Dragonfly and GAMA colors we show color profiles measured by Dragonfly. Here we focus on galaxies with S\'ersic measured $r_{\rm eff, r} > 3\rm \ arcsec$ (compared to the median value of $\sim  1.5\rm \ arcsec$). Since the Dragonfly PSF HWHM is $\sim 2.5$ arcsec, these profiles shown should be well-resolved at at $r \gtrsim r_{\rm eff, r}$, but it is important to keep in mind that these profiles are not de-convolved from the PSF.

\begin{figure}
    \centering
    \includegraphics[width = 0.99 \columnwidth]{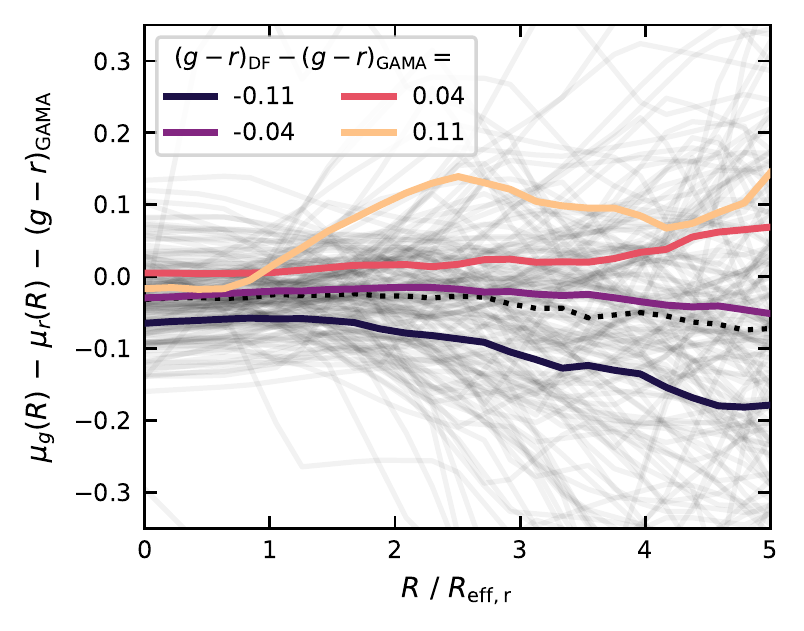}
    \caption{Normalized color profiles measured by Dragonfly. The color profiles for each galaxy are normalized both by their effective radius and its GAMA measured color. Thin grey lines show individual profiles and the dotted line shows the median color profile of the entire sample. The colored lines show median normalized profiles of galaxies binned by the difference between Dragonfly and GAMA measured total color. Galaxies which are bluer in Dragonfly, i.e. have a negative color difference, show a negatively sloped color profile and vice-versa for galaxies which are redder in Dragonfly. For this analysis, note that we only focus on galaxies with $r_{\rm eff, r} > 3\rm \ arcsec$. }
    \label{fig:mu_comp}
\end{figure}

Color profiles measured by Dragonfly are shown in Figure~\ref{fig:mu_comp}. For each galaxy we normalize the color profile by $(g-r)_{\rm \texttt{GAMA} }$ and the radius by $R_{\rm eff, r}$. At small radii, all of the Dragonfly measurements agree with the GAMA colors. At larger radii we see the scatter in individual color profiles grows. We also show the median normalized profile of galaxies binned by the difference in total color, $(g-r)_{\rm DF} - (g-r)_{\rm GAMA}$. Galaxies where the integrated Dragonfly color is bluer (i.e. negative  $(g-r)_{\rm DF} - (g-r)_{\rm \texttt{GAMA}}$ ) generally have negative color gradients at $R > R_{\rm eff.}$. Conversely, galaxies with redder Dragonfly colors have positive color gradients.

The presence of these color gradients, and the correlation with the difference in total color, provides an explanation for the difference between the GAMA colors and the Dragonfly colors. The GAMA colors are measured within the Kron radius of the SDSS images, which is typically 1-1.5 $r_{\rm eff, r}$. This means it does not capture the effect of the color gradient. In massive galaxies these color gradients are generally negative and Dragonfly is better able to capture the full effects of these gradients, therefore Dragonfly colors are on average bluer.

\section{Implications for Stellar Mass estimates}
\label{sec:res_sm}
\subsection{Deriving corrections to GAMA stellar mass estimates}
In this section we investigate the effects that the observational differences discussed in Sections~\ref{sec:res_sersic} and~\ref{sec:res_auto} have on the estimate of the total stellar mass of a galaxy. There are two effects that we will consider which alter the stellar mass estimate based on the Dragonfly observations. The first is the change to the total luminosity, and the second is the change in mass-to-light ratio due to the change in color.

The first effect is relatively straightforward to account for. We simply compare the total flux observed by Dragonfly in the $r$ band to that measured by GAMA. Again, this is the flux contained within $10\, r_{\rm eff}$ of the $r$ band single component S\'ersic model~\citep{Kelvin2012}. The ratio of $r$ band fluxes then directly translates into the ratio of total luminosities.
 
\begin{figure}
    \centering
    \includegraphics[width = 0.99\columnwidth]{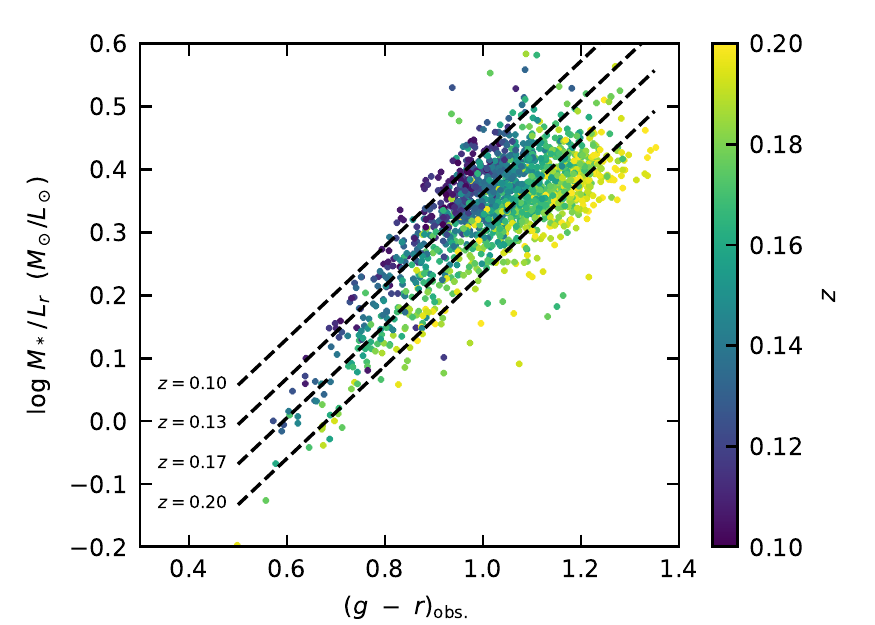}
    \caption{The mass-to-light ratio for our sample of galaxies as measured by~\citet{Taylor2011}, as a function of the observed $g-r$ color. We fit a log-linear relation, with a term for the redshift evolution, to this data shown by the black dashed lines. The best fit parameters are shown in Eqn.~\ref{eqn:MLr}. }
    \label{fig:ML_gmr}
\end{figure}
 
The second effect is more challenging to account for as the mass-to-light ratio is usually calculated using SED fitting of many photometric bands and Dragonfly only measures the $g$ and $r$ band photometry. To approximate this we will assume there is a log-linear relationship between the mass to light ratio and the $g-r$ color. We fit a linear relationship between $ \log M/L_r$ calculated in \citet{Taylor2011} to the $(g-r)_{\rm GAMA}$ color which was used to derive it. Since we are using the observed color we also introduce a redshift term which accounts for the shifting bandpass. The data and fit are shown in Figure~\ref{fig:ML_gmr}. The best fit relation is:
 
 \begin{equation}
     \log M_* / L_r = -0.12\, (1\, +\, 15.7\, z) + 0.74\, (g-r)_{\rm obs.}
     \label{eqn:MLr}
 \end{equation}

From this form of the equation one can show that a difference in mass-to-light ratio depends only on the slope of the equation. Comparing the mass-to-light ratio implied by two different colors, $(g-r)$ and $(g-r)^{\prime}$, we find:

 \begin{equation}
     \frac{ {M_* / L_r}_{(g-r)^\prime} } { {M_* / L_r}_{(g-r)} } = 10^{ 0.74\, \Delta (g-r)}
 \end{equation}

Here, $\Delta (g-r) = (g-r)^\prime - (g-r)$ represents the difference between the colors. Using this equation we can easily compare the mass-to-light ratios inferred for Dragonfly colors and other measurements.

\begin{figure*}
    \centering
    \includegraphics[width = \textwidth]{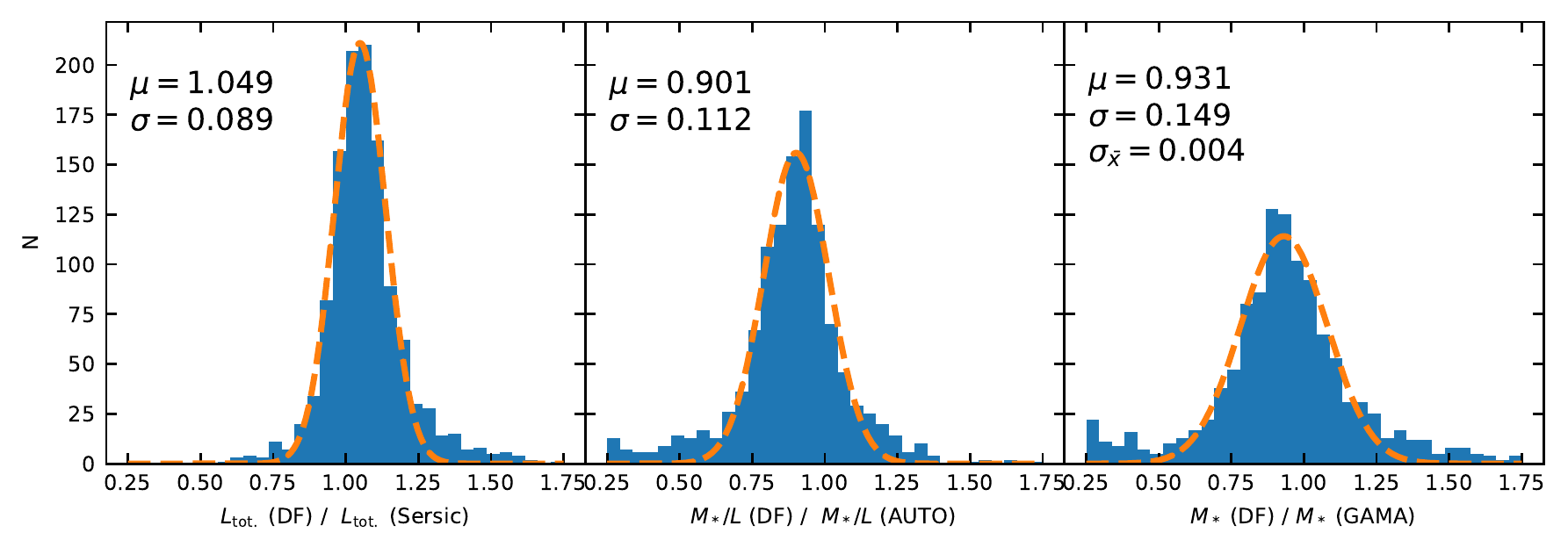}
    \caption{The effect of Dragonfly photometry on the GAMA stellar mass estimates. We show the distribution of differences in total luminosity (left), implied mass-to-light ratio (middle) and the combination of the two effect on the the total stellar mass (right). Dashed lines shown Gaussian fits to each distribution with the parameters listed in each panel.}
    \label{fig:sm_ratio}
\end{figure*}

Figure~\ref{fig:sm_ratio} shows how the Dragonfly measurements affect the estimate of the total stellar mass. We show the distribution of total luminosity and mass-to-light ratios, comparing Dragonfly to those implied by the GAMA measurements. We also show the total effect on the stellar mass measurement when accounting for both effects. The difference in $r$ band magnitude results in the Dragonfly estimate of the total luminosity being 5\% higher, whereas the mass-to-light ratio inferred by Dragonfly is 10\% lower due to the bluer colors. These two results oppose each other, and the total effect is to lower the stellar mass by $7\%$ compared to the methods used by the GAMA survey. In this final panel we display the standard deviation of the $M_* ({\rm DF}) /  M_* ({\rm GAMA})$ distribution; $\sigma = 0.149$. Also shown is the error the mean, $\sigma_{\bar{x}} = \sigma / \sqrt{N}$, where $N$ is the total number of galaxies in our sample. For the $M_* ({\rm DF}) /  M_* ({\rm GAMA})$ distribution, $\sigma_{\bar{x}} = 0.004$.

\begin{figure*}
    \centering
    \includegraphics[width = \textwidth]{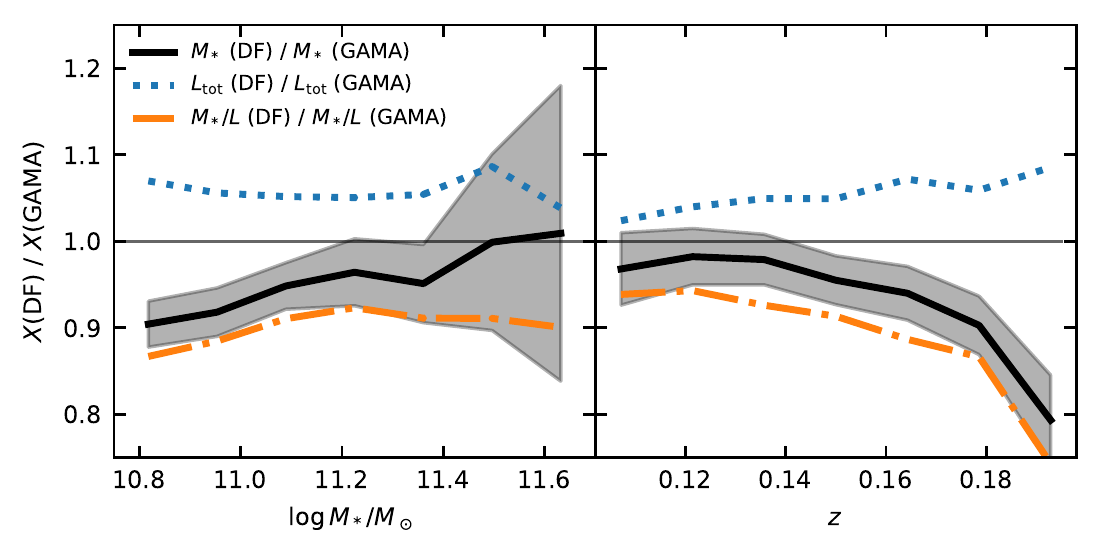}
    \caption{Observational differences between DWFS and GAMA as a function of the GAMA stellar mass (left) and redshift (right). The grey regions show the $2\sigma_{\bar{x}}$ range of the differences of the combined effect on the stellar mass in each stellar mass or redshift bin.}
    \label{fig:sm_phys}
\end{figure*}

The median of these ratios for galaxies in different stellar mass and redshift bins is shown in Figure~\ref{fig:sm_phys}. The median of $M_* ({\rm DF}) /  M_* ({\rm GAMA})$ appears to increase slightly up to $\log M_* / M_\odot \approx 11.2$. At higher masses it appears to remain constant but there are few galaxies in this mass range so we can not confirm this trend with confidence. This is driven mostly by a change in mass-to-light ratio. A possible explanation is that the slope of color gradients, which are responsible for the difference between the GAMA and Dragonfly mass-to-light ratios, vary systematically with stellar mass~\citep{Wang2019,Suess2020}. These ratios also depend on redshift. While the ratio of total luminosities increases slightly with redshift, this change again appears to be mainly due to the difference in mass-to-light ratio, which evolves more rapidly. The reason for this rapid evolution at $z>0.15$ is not immediately clear. One possible explanation is a simple difference in signal-to-noise. At higher redshifts, the surface brightness profiles outside of $r_{\rm eff}$ may be below the noise limit of SDSS. If more of the profile is below the noise limit, the failure to account for color gradients becomes more pronounced, increasing the difference between the GAMA aperature colors and Dragonfly colors.

\subsection{Effect on the measured SMF}
\label{sec:smf}
In this section we investigate how the systematic differences in stellar mass estimates, highlighted in Fig~\ref{fig:sm_ratio} and Fig~\ref{fig:sm_phys}, affect the stellar mass function. Since our analysis sample is incomplete due to a complex series of quality cuts, we can not simply re-measure the SMF using traditional methods and our updated measurements. Instead we calculate the effects implied by the updated photometry using a previously measured SMF. We will be using the double Schecter fit to the bolometric masses found in~\citet{Wright2017} using data from the GAMA survey.\citeauthor{Wright2017} use LAMBDAR photometry in SED fitting to calculate the mass-to-light ratio normalized to the total flux of the r-band S\'ersic model. Using their measured SMF as the probability distribution, we randomly sample a set of $10^7$ galaxies. We then apply a multiplicative ``Dragonfly correction'' to each galaxy's stellar mass based on the results in Figure~\ref{fig:sm_ratio}. For simplicity, we will assume that this correction is independent of stellar mass. We then re-measure the SMF based on the ``corrected'' mass measurements and compare to the original SMF.

We will use two different procedures based on the interpretation of the width of the $\rm M_*(DF)/M_*(GAMA)$ distribution. If the width of this distribution is caused largely by observational errors, then the mean value should be applied as the correction to all galaxies. Conversely, if the width is entirely caused by intrinsic variation within the galaxy population, then it would be correct to apply a different correction to all galaxies. Specifically each correction should be drawn from the distribution shown. In other words, the former is akin to treating the correction as a systematic error whereas the latter is akin to a random error. The truth is likely somewhere in between these two cases. Since the origin of the uncertainties on our total flux measurements are uncertain (see Section~\ref{sec:vtest}) we cannot discriminate between these two scenarios and therefore will consider both as limiting cases.

\begin{figure}
    \centering
    \includegraphics[width = \columnwidth]{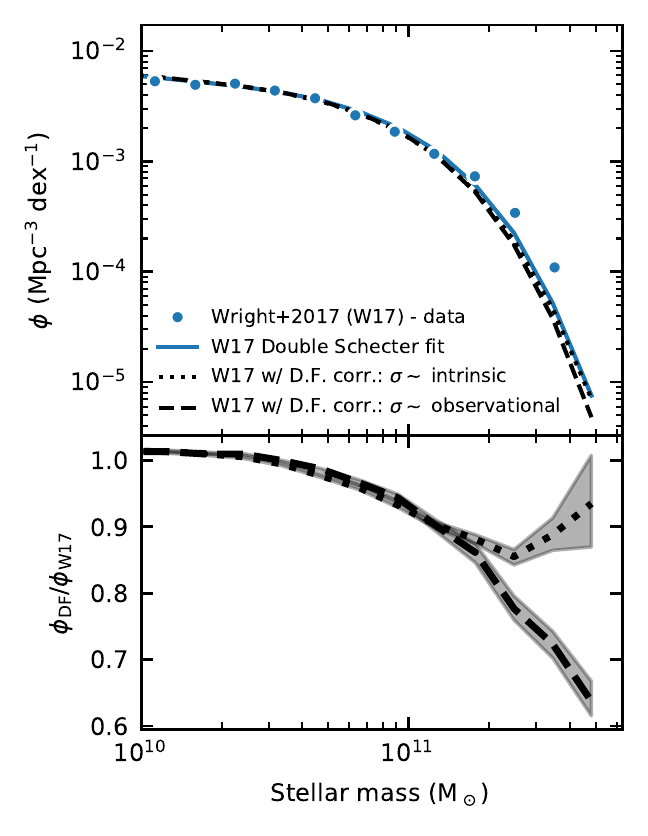}
    \caption{The top panel displays the SMF measured by~\citet{Wright2017} using GAMA data and a modified SMF simulating the effects of Dragonfly like photometry. We do this using two methods: assuming that the width of the $\rm M_*(DF)/M_*(GAMA)$ distribtuion shown in Figure~\ref{fig:sm_ratio} is driven by either intrinsic scatter or observational uncertainty. The bottom panel shows the ratio of the simulated Dragonfly corrected SMF to the original \citeauthor{Wright2017} measurement. The grey region shows the 16th-84th percentile range of $10^4$ bootstrapping samples.}
    \label{fig:SMF_comp}
\end{figure}

In the first scenario we will apply a single correction to all galaxies. This procedure is repeated $10^3$ times, drawing this correction from a Gaussian distribution with $\mu = 0.931$ and $\sigma = 0.004$. This is the error on the mean measured from the distribution of $\rm M_*(DF)/M_*(GAMA)$. In the second case we  apply a different multiplicative correction to each galaxy. These corrections are drawn from a Gaussian distribution with $\mu = 0.931$ and $\sigma = 0.15$, again derived from the results in Figure~\ref{fig:sm_ratio}. This process is also repeated $10^3$ times. For each scenario we show the median and 5\% - 95\% percentile of DF-corrected SMF at a given stellar mass in Figure~\ref{fig:SMF_comp}.

Both procedures show a relatively minor effect on the stellar mass function. At stellar masses less then $10^{11}\ M_\odot$ the effect is less then $5\%$ and and only reaches a maximum of 30\% by $M_* \sim 10^{11.75}\ M_\odot$. This is a relatively minor effect which does not change the overall shape of the SMF significantly. The two limiting cases have slightly different effects with the intrinsic scatter scenario resulting in less difference overall. %However, the two scenarios considered have opposite effects at masses greater the $M_*$. In the first scenario, where single correction that is less then one is applied to all galaxies, the SMF is consistently lower then the original. Whereas in scenario two, which introduces an additional uncertainty, the SMF increases. The median correction is less then one, but due to the exponential decline in the number of galaxies, more galaxies scatter to high masses then scatter to lower. 

We have paramaterized the Dragonfly correction to the \citet{Wright2017} SMF in Appendix~\ref{sec:smf_param}.

\section{Comparison to other methods}
\label{sec:meth_comp}

\begin{figure*}
    \centering
    \includegraphics[width = 0.95\textwidth]{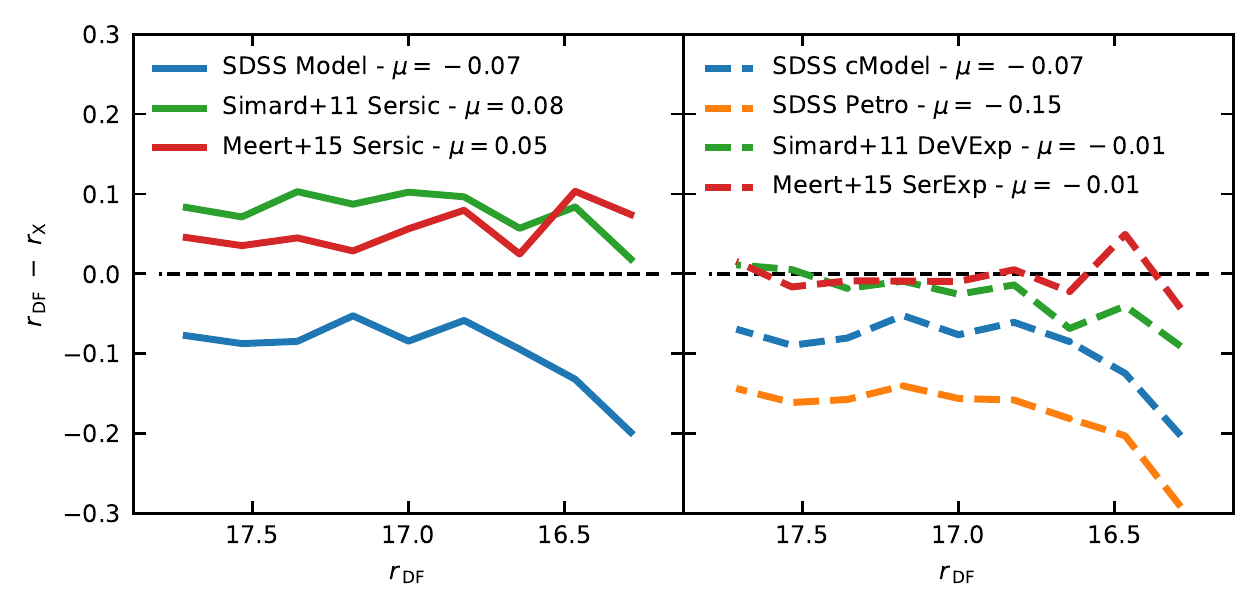}
    \caption{Comparing Dragonfly measured $r$ band magnitudes to various methods performed on SDSS imaging. We compare to \texttt{model}, \texttt{cmodel} and \texttt{Petro} measurements produced by the SDSS pipeline along with S\'ersic or bulge + disk decompositions performed by \citet{Meert2015} and \citet{Simard2011}.}
    \label{fig:obs_comp_mag}
\end{figure*}

As the SDSS data are publicly available many studies have re-analyzed the raw imaging data. We compare our total flux measurements to \citet{Simard2011} and \citet{Meert2015}. Both apply updated sky-subtraction algorithms and apply bulge + disk decompositions or single component S\'ersic models to extract the photometry of galaxies. \citet{Meert2015} parameterize the bulge as a S\'ersic profile, with $n$ as a free parameter and the disk is fixed as an exponential. For the \citet{Simard2011} deVExp measurements, the bulge is fixed as a de Vaucouleurs profile and the disk is fixed as an exponential. The \citet{Meert2015} SerExp photometry measurements are employed as the total flux measurement to calculate the SMF \citet{Bernardi2013} and \citet{Bernardi2017}. The \citet{Simard2011} deVExp measurements are used to calculate the stellar masses used by \citet{Thanjavur2016} to measure the SMF. Additionally we compare to the \texttt{model}, \texttt{cmodel} and \texttt{Petro} measurements from the SDSS photometric catalog~\citet{Ahn2014}. The \texttt{model} photometry is often used across multiple bands when performing SED fitting while the \texttt{cmodel} and \texttt{Petro} measurements have been employed as measurements of total flux\citet{Bell2003,Li2009,Moustakas2013}

Comparisons between their photometry and our Dragonfly photometry is shown in Figure~\ref{fig:obs_comp_mag}. The total magnitudes from the bulge + disk decompositions by \citet{Meert2015} and \citet{Simard2011} both agree well with our measurements. However the magnitude difference between the Dragonfly and \citet{Simard2011} measurements appears to decrease for brighter galaxies. The single component S\'ersic models by both of these studies are brighter on average then the Dragonfly measurements. This is the total flux of the S\'ersic model (i.e. integrated to infinity) which may explain the difference between the GAMA measurements, which are truncated at 10 $r_{\rm eff.}$. The \texttt{model}, \texttt{cmodel} and \texttt{petro} magnitudes reported in the SDSS database severely underestimate the flux measured by Dragonfly by up to 0.3 mag for the brightest galaxies, echoing the results of previous studies ~\citep{Bernardi2013,Dsouza2015}.

\begin{figure}
    \centering
    \includegraphics[width = 0.95\columnwidth]{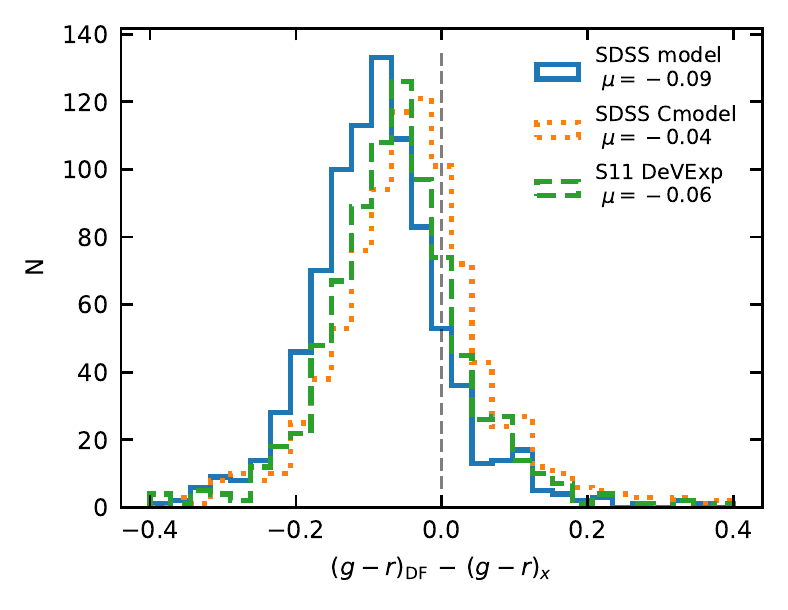}
    \caption{Dragonfly colors compared to \texttt{model}, \texttt{cmodel} and deVExp measurements performed by \citet{Simard2011}. Similar to the aperture matched measurements, Dragonfly measures bluer colors then these other methods based on SDSS imaging.}
    \label{fig:obs_comp_col}
\end{figure}

In addition to aperture photometry, the \texttt{model} and \texttt{cmodel} photometry provided by the SDSS pipeline is also commonly used in SED fitting to calculate stellar masses~\citep{Li2009,Moustakas2013,Bernardi2013}. \citet{Mendel2014} use the \citet{Simard2011} deVExp decompositions in their SED fittings. Figure~\ref{fig:obs_comp_col} shows how these methods compare to our Dragonfly color measurements. Similar to the GAMA measurements, the Dragonfly $g-r$ are on average bluer then these SDSS measurements. The mean color difference is -0.09 mag and -0.04 mag when comparing the Dragonfly colors to the \texttt{model} \texttt{cmodel} colors, respectively. The \citet{Simard2011} two component decompositions also show bluer colors on average the Dragonfly, with the mean color difference being -0.06 mag. Unlike the \texttt{AUTO} method employed by GAMA these are not aperture measurements, therefore the cause of this discrepancy cannot be understood as simply as with color gradients described above.

\section{Discussion} \label{sec:disc}
In this study we present photometry of galaxies measured in the DWFS and compare to previous methods for measuring the photometry of massive galaxies at $\log M_*/ M_\odot > 10.75$. Our aim was to develop a flexible and non-parametric method which takes advantage of Dragonfly's low surface-brightness sensitivity. We then compare the total flux and color measurements to results from GAMA, derived from SDSS imaging, and other methods which use the same imaging dataset. Our measurements provide an independent test of the photometric methods which are currently used for massive galaxies.

Perhaps the most interesting result of this study is that the S\'ersic or SerExp models performed by \citet{Simard2011},\citet{Kelvin2012} and \citet{Meert2015} generally match our Dragonfly total flux measurements well. This mirrors previous results from studies with Dragonfly ~\citep{vanDokkum2014,Merritt2016} which often find a smaller then expected stellar halo around nearby spiral galaxies, the so called missing outskirts problem~\citep{Merritt2020}. On average we measure profiles out to $10\, r_{\rm eff}$, or roughly 15\%-20\% of $r_{\rm vir}$, and down to $\mu_g \approx 31\ \rm mag\ arcsec^{-2}$. Yet, we find little evidence for significant variation ($\gtrsim 5\%$) from a S\'ersic or S\'ersic-Exponential profile. Given simulations and empirical models suggest a large amount of mass in the IHL~\citep{Pillepich2018,Behroozi2019}, this is perhaps surprising. In a recent study~\citet{Merritt2020} suggest that the TNG100 cosmological simulation over-predict the amount of light in stellar halos of Milky Way mass galaxies. In future studies we plan to study the profile of individual galaxies and compare to theoretical models predicting the amount and extent of the diffuse stellar halo or IHL. 

Our method of integrating the 1-D surface brightness profile is similar to those applied by \citet{Huang2018} and \citet{Wang2019} to Hyper-Suprime Cam (HSC) imaging of massive galaxies. \citeauthor{Huang2018} measure the profiles of individual massive galaxies at $z\sim0.4$ galaxies out to, and beyond, 100 kpc. By integrating these profiles, the authors calculate the total flux. They find that using over simplified assumptions or shallow imaging misses a significant fraction ($\gtrsim 20 \%$) of the total light. Moreover, this discrepancy depends both on the stellar mass and halo mass of the galaxy. \citet{Wang2019} perform a stacking analysis to study the stellar halos of isolated central galaxies. Applying a similar method of integrating the 1-D profile, they find that the \texttt{cmodel} method underestimates the total light of galaxies at $10 < \log M_* / M_\odot < 11.5$ by $\lesssim 10\%$ and interestingly that $\sim 10 \%$ of the total light is beyond the noise limit of a single (non-stacked) HSC image. Our results generally agree with both of these studies that the \texttt{cmodel} method underestimate the total light by up to $20\%$. 

Our results are also consistent with~\citet{Bernardi2017}, who conclude that different methods of calculating total flux only alter the SMF at the level of $\sim 0.1$ dex (or $\sim 20\%$). This is secondary, the authors argue, to differences caused by different treatments and assumptions of stellar populations used in SED fitting that result in systematic variations in the SMF of order $\lesssim 0.5$ dex. It is important to clarify that this is a separate issue from what we discuss in this study. \citeauthor{Bernardi2017} discuss how different methods derive different mass-to-light ratios from, mostly, the same photometry, whereas we focus on how systematically biased photometry can alter the implied mass-to-light ratio. Along with the measurement of total flux, the accurate measurement of colors across multiple bands adds an additional systemic issue to the stellar mass estimates of massive galaxies. Comparing to different established techniques, Dragonfly consistently measures bluer colors implying a lower mass-to-light ratio and therefore lower stellar mass.

We show that the discrepancy between the Dragonfly and aperture measured colors is caused by color gradients. This is an inherent shortcoming of the aperture photometry technique. The bluer colors measured by Dragonfly also corresponds with our understanding about the redder colors of bulges compared to disks~\citep{Lackner2012} and the general trend of negative color gradients in massive galaxies~\citep{Saglia2000,LaBarbera2005,Tortora2010}. \citet{Huang2018} and \citet{Wang2019} also observe negative color gradients in massive galaxies out to $\sim 50$ kpc. \citeauthor{Wang2019} find some evidence of an upturn in the color profile at larger radii, however the authors note this result is sensitive to the details of PSF deconvolution and masking of nearby sources. They additionally find that the gradient becomes shallower for more massive galaxies.

On-going surveys such as \textit{DECaLs}~\citep{Dey2019}, \textit{DES}\citep{DES2018}, HSC-SSP~\citep{Aihara2019} and the upcoming Rubin Observatory~\citep{Ivezic2019} are set to provide higher quality images over a comparably large area as SDSS. It is unclear if current methods applied to this data will produce more accurate results. \citet{Huang2018} and \citet{Wang2019} compare their non-parametric measurements to the \texttt{cmodel} method applied to HSC images. They find the HSC \texttt{cmodel} magnitudes underestimate the total flux by $10\% \text{--} 25\%$. This is likely due to the rigidity of the \texttt{cmodel} paramaterization which does not match the surface brightness profiles of massive galaxies. In order to implement a non-parametric method, these surveys will need to measure galaxy surface brightness profiles down to $\mu_r \lesssim 30.5\ \rm mag\ arcsec^{-2}$, we find that on average for 99\% of the total flux of the galaxy is brighter than this limit. Additionally deeper data will allow the color to be measured within a larger aperture, limiting the effect of color gradients. For example,  \citep{Bellstedt2020} calculate the flux using a curve of growth and define convergence when the flux changes by $<5\%$, which might lead to uncertainties on the order of 5\%. An additional issue with deeper data is that there is more contamination from neighboring objects, making de-blending more of a challenge

While we designed our method to be as non-parametric as possible, some extrapolation is still necessary. Injection-recovery tests performed in Section~\ref{sec:method} show that our method accurately recovers the total flux of S\'ersic like profiles, however this does not necessarily reflect the truth. If there is an over (under) abundance of light below our noise limit of 31 mag arcsec$^{-2}$ compared to these profiles, we could be systematically under (over) predicting the total flux of these galaxies. Currently, the only reliable method to probe below this limit is individual star counts of nearby halos, however this is very resource intensive~\citep{Radburn-Smith2011}. Using data from the GHOSTS survey, \citet{Harmsen2017} show that the minor axis profiles of six nearby disk galaxies are consistent with a power-law of slope $-2$ to $-3.7$, but with a large amount of intrinsic scatter consistent with \citet{Merritt2016} and \citet{Merritt2020}. However, this is not a perfect comparison. With extremely deep exposures ($\gtrsim 100$ hr) or stacking techniques it may be possible to reach below 31 mag arcsec$^{-2}$ with Dragonfly, but this too will require significant effort.

\section{Summary} \label{sec:conc}
In this work we present measurements of the $g$ and $r$ band photometry of massive galaxies in the DWFS. We focus on galaxies with $\log M_* / M_\odot > 10.75$ at $0.1 <z< 0.2$. To take advantage of the low surface brightness sensitivity of Dragonfly, we develop a method for measuring photometry based on integrating the 1-D surface brightness profile that is minimally dependent on any paramaterization. The catalog of photometry measurements for the final analysis sample is available here\footnote{\url{https://tbmiller-astro.github.io/data/}}. We then compare our measurements to various methods applied to SDSS imaging, focusing on those favoured by the GAMA survey. In particular, we focus on the $r$ band total flux and $g-r$ color and their implications for the stellar mass estimates of massive galaxies. Our main results are summarized below:

\begin{itemize}
    \item First we compare the Dragonfly $r$ band total flux to S\'ersic models measured by \citet{Kelvin2012}. We find the Dragonfly measurements are brighter by $0.05 \pm 0.09$ mag compared to their S\'ersic model, truncated at $10\ r_{\rm eff}$. When comparing to other methods, we find that the SDSS reported \texttt{model}, Petro and \texttt{cmodel} measuresments severely underestimate the total flux by up to 0.3 mag, echoing previous results~\citep{Bernardi2013,Dsouza2015}. Additionally, the bulge + disk decompositions performed by \citet{Meert2015} and \citet{Simard2011} match our Dragonfly measurements well.
    \item Comparing the Dragonfly $g-r$ colors to the aperature photometry performed by \citet{Wright2016} we find on average bluer colors by $0.06 \pm 0.07$ mag. By measuring the color profile we show that this discrepancy is caused by color gradients which are generally negative for massive galaxies (Fig. \ref{fig:mu_comp}). Comparing to other established methods, such as \texttt{model}, \texttt{cmodel} and \citet{Simard2011} decompositions, we find the Dragonfly measured colors are bluer by an average of 0.09 mag, 0.04 mag and 0.06 mag, respectively.
    \item When considering the effect on GAMA stellar mass estimate, the two discrepancies discussed above have opposing effects. The larger total flux measured by Dragonfly implies an average of $5\%$ larger total luminosity but the bluer colors results in $10\%$ lower mass-to-light ratio on average. The combined effect is that the stellar mass estimate is $7\%$ lower when accounting for the difference between GAMA and Dragonfly photometry. 
    \item Finally, we estimate the effect these corrections will have on the measured SMF. Comparing to the \citet{Wright2017} SMF, we find a relatively small difference, $\lesssim 30\%$, in number density of galaxies at $\log M_*/M_\odot = 11.5$.
\end{itemize}

When comparing to SDSS data, we find that multi-component 2-D decompositions are the most accurate way to measure the photometry of nearby massive galaxies. However, these methods remain very computationally expensive. With higher quality data it may be possible to employ other, less resource intensive or non-parametric, methods that provide similarly accurate results. For massive galaxies this will rely on the low surface brightness sensitivity and accurate sky modelling to measure the extended light profiles. Dragonfly measurements will remain an important benchmark for future surveys and methods.

\acknowledgments
T.M. would like to thank Patricia Gruber and the Gruber foundation for their generous support of the work presented here. S.D. is supported by NASA through Hubble Fellowship grant \# HST-HF2-51454.001-A awarded by the Space Telescope Science Institute, which is operated by the Association of Universities for Research in Astronomy, Incorporated, under NASA contract NAS5-26555. J.P.G. is supported by an NSF Astronomy and Astrophysics Postdoctoral Fellowship under award AST-1801921. The authors thank the excellent and dedicated staff at the New Mexico Skies Observatory. We also thank S\'{e}bastien Fabbro for his support and help with the CANFAR services. Support from NSF grant AST1613582 is gratefully acknowledged. 

GAMA is a joint European-Australasian project based around a spectroscopic campaign using the Anglo-Australian Telescope. The GAMA input catalogue is based on data taken from the Sloan Digital Sky Survey and the UKIRT Infrared Deep Sky Survey. Complementary imaging of the GAMA regions is being obtained by a number of independent survey programmes including GALEX MIS, VST KiDS, VISTA VIKING, WISE, Herschel-ATLAS, GMRT and ASKAP providing UV to radio coverage. GAMA is funded by the STFC (UK), the ARC (Australia), the AAO, and the participating institutions. The GAMA website is http://www.gama-survey.org/.

The Legacy Surveys consist of three individual and complementary projects: the Dark Energy Camera Legacy Survey (DECaLS; NOAO Proposal ID \# 2014B-0404; PIs: David Schlegel and Arjun Dey), the Beijing-Arizona Sky Survey (BASS; NOAO Proposal ID \# 2015A-0801; PIs: Zhou Xu and Xiaohui Fan), and the Mayall z-band Legacy Survey (MzLS; NOAO Proposal ID \# 2016A-0453; PI: Arjun Dey). DECaLS, BASS and MzLS together include data obtained, respectively, at the Blanco telescope, Cerro Tololo Inter-American Observatory, National Optical Astronomy Observatory (NOAO); the Bok telescope, Steward Observatory, University of Arizona; and the Mayall telescope, Kitt Peak National Observatory, NOAO. The Legacy Surveys project is honored to be permitted to conduct astronomical research on Iolkam Du'ag (Kitt Peak), a mountain with particular significance to the Tohono O'odham Nation.

NOAO is operated by the Association of Universities for Research in Astronomy (AURA) under a cooperative agreement with the National Science Foundation.

This project used data obtained with the Dark Energy Camera (DECam), which was constructed by the Dark Energy Survey (DES) collaboration. Funding for the DES Projects has been provided by the U.S. Department of Energy, the U.S. National Science Foundation, the Ministry of Science and Education of Spain, the Science and Technology Facilities Council of the United Kingdom, the Higher Education Funding Council for England, the National Center for Supercomputing Applications at the University of Illinois at Urbana-Champaign, the Kavli Institute of Cosmological Physics at the University of Chicago, Center for Cosmology and Astro-Particle Physics at the Ohio State University, the Mitchell Institute for Fundamental Physics and Astronomy at Texas A\&M University, Financiadora de Estudos e Projetos, Fundacao Carlos Chagas Filho de Amparo, Financiadora de Estudos e Projetos, Fundacao Carlos Chagas Filho de Amparo a Pesquisa do Estado do Rio de Janeiro, Conselho Nacional de Desenvolvimento Cientifico e Tecnologico and the Ministerio da Ciencia, Tecnologia e Inovacao, the Deutsche Forschungsgemeinschaft and the Collaborating Institutions in the Dark Energy Survey. The Collaborating Institutions are Argonne National Laboratory, the University of California at Santa Cruz, the University of Cambridge, Centro de Investigaciones Energeticas, Medioambientales y Tecnologicas-Madrid, the University of Chicago, University College London, the DES-Brazil Consortium, the University of Edinburgh, the Eidgenossische Technische Hochschule (ETH) Zurich, Fermi National Accelerator Laboratory, the University of Illinois at Urbana-Champaign, the Institut de Ciencies de l'Espai (IEEC/CSIC), the Institut de Fisica d'Altes Energies, Lawrence Berkeley National Laboratory, the Ludwig-Maximilians Universitat Munchen and the associated Excellence Cluster Universe, the University of Michigan, the National Optical Astronomy Observatory, the University of Nottingham, the Ohio State University, the University of Pennsylvania, the University of Portsmouth, SLAC National Accelerator Laboratory, Stanford University, the University of Sussex, and Texas A\&M University.

BASS is a key project of the Telescope Access Program (TAP), which has been funded by the National Astronomical Observatories of China, the Chinese Academy of Sciences (the Strategic Priority Research Program "The Emergence of Cosmological Structures" Grant \# XDB09000000), and the Special Fund for Astronomy from the Ministry of Finance. The BASS is also supported by the External Cooperation Program of Chinese Academy of Sciences (Grant \# 114A11KYSB20160057), and Chinese National Natural Science Foundation (Grant \# 11433005).

The Legacy Survey team makes use of data products from the Near-Earth Object Wide-field Infrared Survey Explorer (NEOWISE), which is a project of the Jet Propulsion Laboratory/California Institute of Technology. NEOWISE is funded by the National Aeronautics and Space Administration.

The Legacy Surveys imaging of the DESI footprint is supported by the Director, Office of Science, Office of High Energy Physics of the U.S. Department of Energy under Contract No. DE-AC02-05CH1123, by the National Energy Research Scientific Computing Center, a DOE Office of Science User Facility under the same contract; and by the U.S. National Science Foundation, Division of Astronomical Sciences under Contract No. AST-0950945 to NOAO.

\software{Astropy~\citep{Astropy2018}, Photutils~\citep{photutils} SEP~\citep{Bertin1996,SEP}, Galsim~\citep{Galsim}, mrf~\citep{vanDokkum2019a}, astroquery~\citep{astroquery}, numpy~\citep{numpy}, scipy~\citep{scipy}, matplotlib~\citep{matplotlib}, pandas~\citep{pandas} }

\newpage
\bibliographystyle{aasjournal}
\bibliography{DFWFS_galphot,software}{}

%\facilities{HST(STIS), Swift(XRT and UVOT), AAVSO, CTIO:1.3m, CTIO:1.5m,CXO}

\newpage
\appendix
\section{Deriving a conversion between Dragonfly and SDSS Filters}
\label{sec:filt_conv}
Even though Dragonfly uses very similar filter sets to SDSS, there are slight differences in the throughput due mostly to differences in quantum efficiencies of the detectors~\citep{Abraham2014}. For a unbiased comparison between the two surveys we must derive a conversion between the two filter sets. To derive this relationship, we implement and compare two commonly used methods: synthetic photometry~\citep{Jester2005, Jordi2010} and an empirical calibration based on the photometry of standard stars~\citep{Jordi2006}.

For the synthetic photometry, we employ the python package \texttt{sedpy}\footnote{https://github.com/bd-j/sedpy} using the spectra of standard stars from the Gunn-Stryker atlas~\citep{Gunn1983} and published filter transmission curves for Dragonfly~\citep{Abraham2014} and SDSS~\citep{Doi2010}. We then fit to the magnitude difference between the SDSS and Dragonfly filters, $m_{\rm SDSS} - m_{\rm DF}$ as a simple linear function of the Dragonfly color, $g_{\rm DF} - r_{\rm DF}$. The results for the $g$ and $r$ band are shown below. The r.m.s of each fit is $\sim 0.003$ mag.
\begin{equation}
    g_{\rm SDSS} - g_{\rm DF} = 0.0965 (g_{\rm DF} - r_{\rm DF}) - 0.0055 
    \label{eqn:synphot_filt_g}
\end{equation}

\begin{equation}
r_{\rm SDSS} - r_{\rm DF} = 0.0331 (g_{\rm DF} - r_{\rm DF}) + 0.0025
\label{eqn:synphot_filt_r}
\end{equation}

For the empirical calibration we use the star catalog from the GAMA \texttt{SpStandards} with measurements from SDSS DR7. We then measure the photometry of these stars in the DWFS frames using the same method we use to measure the total luminosity of galaxies, described in Sec.~\ref{sec:method}. From a simple comparison of SDSS and DF measurements we find and r.m.s. of the DF magnitudes of $\sim 0.05$ mag, while the SDSS magnitudes have a typical reported uncertainty of $\sim 0.01$ mag.

For the empirical calibration, it is dangerous to fit simple linear model as with the synthetic photometry. Given the uncertainties of our measurements are comparable to the trend over the observed color range ($0.2 < g-r < 0.6$) and the fact that $m_{\rm SDSS} - m_{\rm DF}$ is not independent from $g_{\rm DF} - r_{\rm DF}$, performing a simple linear fit is not statistically sound. To circumvent these issues we include a third dataset to mediate between the SDSS and Dragonfly measurements. The dataset we use is GAIA DR2~\citep{Gaia2018}, specifically the mean source photometry from the blue photometer, $G_{BP}$, and red photometer, $G_{RP}$~\citep{Riello2018}. We crossmatch the stars between the two catalogs based on sky position.

First we fit the GAIA color as a linear function of the Dragonfly color, $g_{\rm DF} - r_{\rm DF}$ using a simple linear relation shown here: 
\begin{equation}
    G_{RP} - G_{BP} = \beta (g_{\rm DF} - r_{\rm DF}) + \alpha
    \label{eqn:gaia_df}
\end{equation}
Here, $\alpha$ and $\beta$ are free parameters to be fit. Next we fit $x_{\rm SDSS} - x_{\rm DF}$ as a linear function of the GAIA color: 
\begin{equation}
    x_{\rm SDSS} - x_{\rm DF} = b_x (G_{BP} - G_{RP}) + a_x
    \label{eqn:gaia_sdss}
\end{equation}

Here, $x$ represents either the $g$ or $r$ band. Once these two relations are fit using a simple least squares minimization, the results are combined to obtain $ x_{\rm SDSS} - x_{\rm DF}$ as a function of $g_{\rm DF} - r_{\rm DF}$, where the new slope is $b^*_x = b_x\, \beta$ and the new intercept is $a^*_x = a_x\,  + b_x\, \alpha$.

\begin{figure*}
    \centering
    \includegraphics[width = \textwidth]{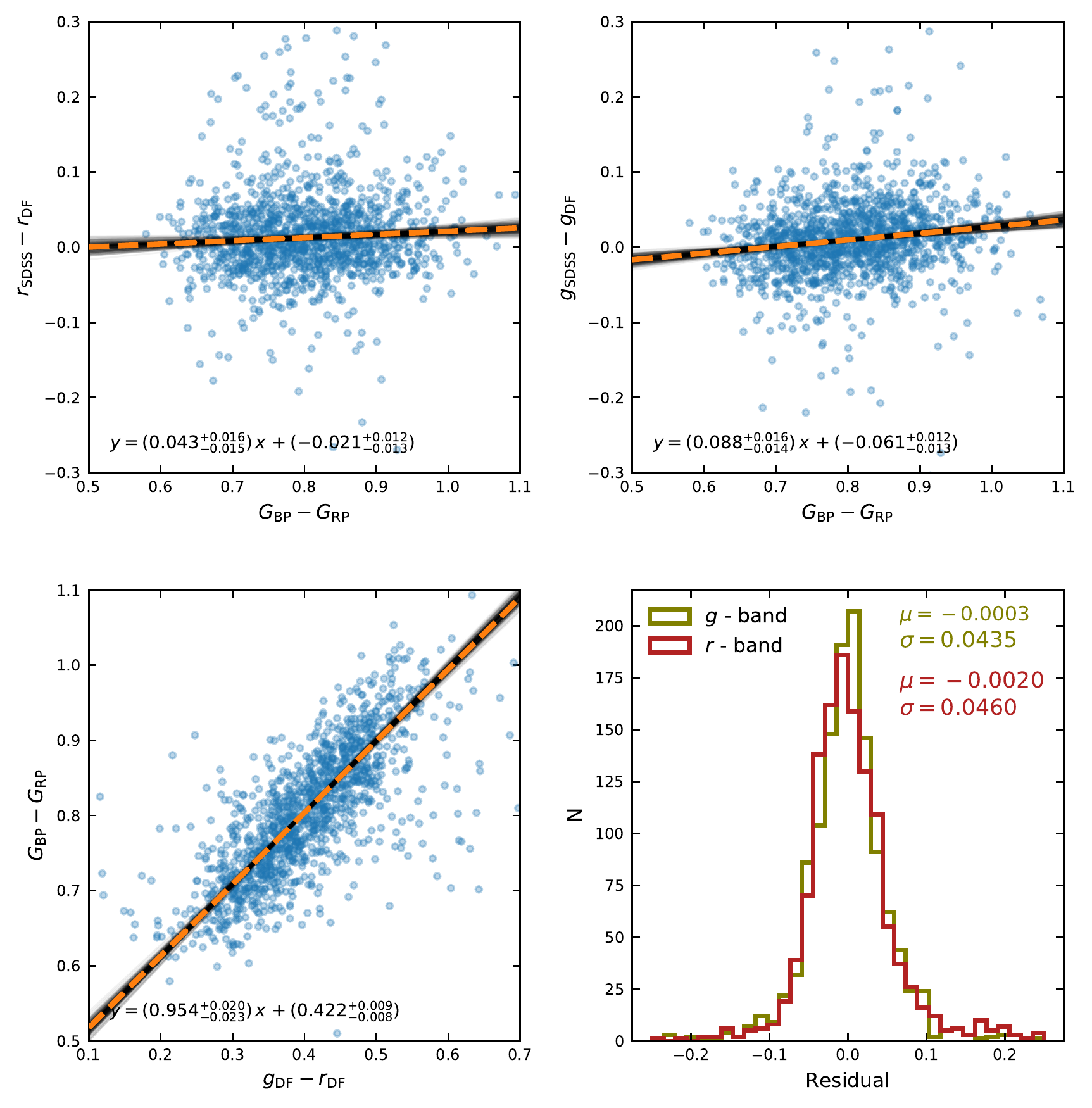}
    \caption{\textbf{Top left:} We show the difference between the SDSS and Dragonfly $r$ band magnitudes for the sample of standard stars as a function of the GAIA color. The thin grey lines show best fit relations for 200 bootstrap samples with the dotted line being the median best fit parameters of $10^5$ bootstrap samples. These parameters are displayed in the bottom left of the panel along with the 16th - 84th percentile range. \textbf{Top right:} Same as left except displaying the difference in measured $g$ band magnitudes. \textbf{Bottom left:} Similar to above except we show the GAIA color as a function of the  Dragonfly color. \textbf{Bottom right:} We show the distribution residuals between the SDSS and Dragonfly measured magnitudes after the filter correction has been applied. For both bands the distribution is roughly gaussian with mean near zero and with of $\sim 0.045$ mag.}
    \label{fig:filt_conv}
\end{figure*}

The results of each step of the empirical calibration are shown in Figure~\ref{fig:filt_conv}. To estimate uncertainties on the parameters we perform a bootstrapping analysis, re-sampling the stars with replacement, for $10^5$ iterations. we take the median of all the parameters as the best fit value and half of the 16th - 84th percentile difference as the $1 \sigma$ uncertainty. The results are shown in the equation below:

\begin{equation}
g_{\rm SDSS} - g_{\rm DF} = (0.087 \pm 0.014)\, (g_{\rm DF} - r_{\rm DF}) + (- 0.024 \pm 0.006)
\label{eqn:emp_filt_g}
\end{equation}

\begin{equation}
    r_{\rm SDSS} - r_{\rm DF} = (0.018 \pm 0.015)\,  (g_{\rm DF} - r_{\rm DF}) + (0.097 \pm 0.006)
\label{eqn:emp_filt_r}
\end{equation}

While there are some disagreements, the two methods are generally consistent. Specifically the slopes for both the $g$ and $r$ band conversions agree within $ 1\sigma$. The intercepts do not match entirely between the synthetic photometry and empirical calibration. Minor differences with zero-point calibration or total throughput measurements of the filter curves could cause this discrepancy. For this reason, we decide to use the empirically calibrated relation for this study. Uncertainties in these derived parameters introduces additional uncertainty into our measurements; $\sim 0.01$ mag for blue galaxies ($g-r \approx 0.6$) and $\sim 0.02$ mag for redder galaxies ($g-r \approx 1.1$).

\vspace{5mm}

\section{Paramaterization of the Dragonfly SMF correction}
\label{sec:smf_param}
Here we paramaterize the difference between the Dragonfly ``corrected" SMF and the original \citet{Wright2017} measurements. We fit a polynomial to the SMF ratio to both limiting cases we consider. We focus on masses $\log M_*/ M_\odot > 10.5$. For the case where the scatter is intrinsic, we fit a third degree polynomial, the best fit parameters are:
\begin{equation}
    \frac{\phi_{\rm DF}\,( \sigma \sim {\rm intr})} {\phi_{\rm W17} } = 0.4\, (\log  M_*/ M_\odot - 10.5)^3 - 0.52\, (\log  M_*/ M_\odot - 10.5)^2 + 1
    \label{eqn:int_smf_corr}
\end{equation}

For the other limiting case, where the scatter in $M_{\rm *,DF} / M_{\rm *,GAMA}$ is driven by observational uncertainty, we fit a second degree polynomial, with the best fit parameters shown here, 
\begin{equation}
    \frac{\phi_{\rm DF}\,( \sigma \sim {\rm obs})} {\phi_{\rm W17} } = -0.26\, (\log  M_*/ M_\odot - 10.5)^2 + 1
    \label{eqn:obs_smf_corr}
\end{equation}

\begin{figure}[!hb]
    \centering
    \includegraphics[width = 0.6\textwidth]{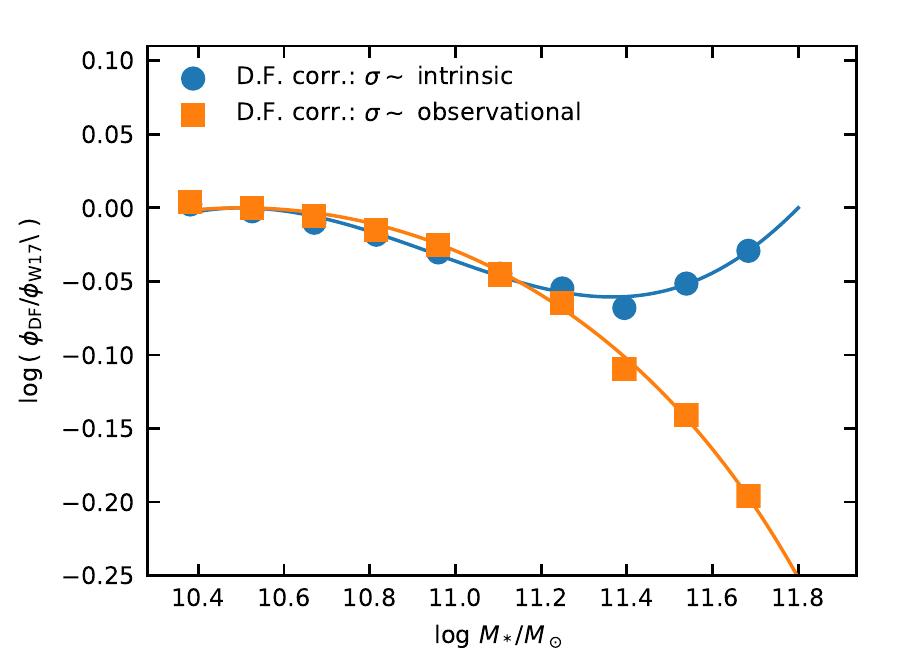}
    \caption{We show the binned ratio of the original~\citet{Wright2017} SMF to the simulated Dragonfly corrected measurements. The two limiting cases discussed in~\ref{sec:smf} are shown. The lines show the best fit polynomials to each set of data points. The best fit relations are displayed in Eqn.~\ref{eqn:int_smf_corr} and  Eqn.~\ref{eqn:obs_smf_corr}}
    \label{fig:smf_corr}
\end{figure}

The data and these best fit relationships are shown in Figure~\ref{fig:smf_corr}. We wish to emphasize that these corrections are focused on this specific form of the SMF and $\log M_*/ M_\odot > 10.5$. Specifically we caution against using this at lower masses as the polynomial fits diverge quickly. We find little change in the SMF at lower masses as the SMF is much flatter then at high masses. Additionally this is derived based on the differences between Dragonfly and GAMA photometry, different surveys that use different photometric methods will result in different corrections.

\end{document}